\documentclass{jpp}
\usepackage{graphicx}
\usepackage{hyperref}
\usepackage[utf8]{inputenc}
\usepackage[T1]{fontenc}
\usepackage{amsmath}
\usepackage{amssymb}
\usepackage{xcolor}

\shorttitle{Kinetic ITG  localisation}
\shortauthor{E. Rodríguez and A. Zocco}

\title{The Kinetic Ion-Temperature-Gradient-driven instability and its localisation}

\author{E. Rodríguez, A. Zocco}

\affiliation{Max Planck Institute for Plasma Physics, 17491 Greifswald, Germany}

\begin{document}

\maketitle

\begin{abstract}
We construct a description of Ion Temperature Gradient (ITG) driven localised {linear} modes which retains both wave-particle and magnetic drift resonant effects while capturing the field-line dependence of the electrostatic potential. We exploit the smallness of the magnetic drift and the strong localisation of the mode to resolve the problem with a polynomial-gaussian expansion in the field-following co-ordinate. A simple semi-analytical formula for the spectrum of the mode is shown to capture long wavelength Landau damping, ion-scale Larmor radius stabilization, weakening of Larmor radius effects at short-wavelengths and magnetic-drift resonant stabilisation. These elements lead to {linear} spectra with multiple maxima as observed in gyrokinetic simulations in stellarators. Connections to the transition to extended eigenfunctions and those localized by less unfavourable curvature regions (hopping solutions) are also made. The model provides a clear qualitative framework with which to interpret numerically simulated ITG modes {linear} spectra with realistic geometries, despite its limitations for exact quantitative predictions.

\end{abstract}

\section{Introduction}
The ion-temperature-gradient (ITG) is one of the primary  modes driving turbulence in magnetic confinement fusion devices \citep{Rudakov,coppi_rosen_sag,Terry_And_Horton}. As such, much literature exists devoted to understanding this type of instability. The most basic understanding of this mode can be gained by studying the linear response of the system, as described by the linearised gyrokinetic equation \citep{connor1980stability,romanelli} used to evaluate quasineutrality.  Extending our understanding of the mode and its response to the geometry is particularly important in the context of \textit{stellarator} physics.
\par
Linear studies of the ITG driven mode, either analytical or numerical, are numerous in the literature, but they are all fundamentally traceable  to the review of  \cite{leontovichV5}.  Analytical progress often requires some  simplifying assumptions. These occur at two levels. First, by considering   separately the various destabilising mechanisms in the problem. Second, by reducing the complexity of the magnetic field along field lines (e.g., curvature, flux compression or the magnetic field magnitude), often described as constant. Our knowledge of the ITG builds on consideration of special cases that respond to different approximations of the first kind \citep{zocco_xanthopoulos_doerk_connor_helander_2018}. When the bad curvature, together with the temperature gradient, drives the mode unstable, we have a \textit{toroidal} ITG \citep{Terry_And_Horton}. {When destabilisation does not require curvature, but is enabled by the propagation of density along the field lines, with a specific relative phase with respect to temperature} \citep{steveITG}, we have a \textit{slab} ITG \cite{Rudakov, leontovichV5}. While any simplified perspective helps shedding light on the physics of the ITG mode, the selective treatment of the physics can hinder their scope. A pressing case of this is the overlook of the mode localisation along the field line in the \textit{toroidal} branch \citep{Terry_And_Horton}, where the presence of bad curvature is paramount. Magnetic fields generally exhibit alternating regions of good and bad curvature every \textit{connection length}.
% , thus alternating sign,  or when modes are particularly extended \citep{zocco_xanthopoulos_doerk_connor_helander_2018,zocco_podavini,podavini2023electrostatic}.  
\par
A simple theoretical description of the inhomogeneity along the magnetic field line, and the { consequent} behaviour of the ITG mode, is approachable through a \textit{fluid description} \citep{horton1981toroidal,hahm1988properties, romanelli,plunk2014}. Assuming particle parallel streaming to be small (slow ion transit time), a strong drive and small curvature drift, one can describe the ITG through a second order ordinary differential equation along the field line. Such description incorporates key physics ingredients to the problem, and crucially couples the behaviour of the instability to its structure along the field line.
However, while localisation seems so important for a mathematical characterisation of the fluid branches \citep{wesson2011tokamaks,zocco2016} and interpretation of numerical results, its analytical treatment becomes increasingly intricate when Landau damping and the magnetic drift resonance are included. { Without a consistent inclusion of these kinetic elements, the fluid description suffers from fundamental shortcomings in describing the behaviour of microinstabilities at small scales (either large perpendicular wavenumber or small parallel scale). This limitation can lead to wrong expectations in the mode response to changes in geometry. But the latter being of crucial importance, especially in the context of current activities such as stellarator optimisation, we are in need to integrate all elements together.}
% In the review of \cite{leontovichV5}, the authors, while proposing a definitive and complete resonant theory of the slab branch with full Larmor radius effects, treat the field-line localisation in a somewhat heuristic manner.
While much progress has been made in this direction for the development of the theory of Geodesic Acoustic Modes \citep{FulvioZonca_1996,SugWatGAM,SUGAMA_WATANABE_2006,QIU_2018}, a manageable kinetic theory that retains resonant effects and { includes} geometric localisation is desirable. We address this problem in this work.

% In particular, consider a toroidal like ITG mode %localised to a region of strong bad curvature, %bounded by good curvature. A mode is happily %localised in this region. As we squeeze the mode, %the fluid approach predicts that the mode's growth %rate will eventually increase without bound. This %is of course non-physical, and is a result of the %transit time being intrinsically ordered in the %fluid approach; or said otherwise, the fluid model %lacking any kinetic effect, such us the potentially %stabilising Landau damping. A symptom that such %kinetic effects are lacking, and that such a lack %is crucial, is that in the fluid description there %exist a hierarchy of modes with even finer features %along the field line, which within the fluid %picture have ever larger growth rates. Once again, %to regularise these unphysical treats, we must %somehow rescue key kinetic elements. 

\par
In this article we introduce a formal approach to the gyrokinetic equation to describe localised ITG modes and its structure in a simple geometry with good and bad curvature alongside kinetic effects, both resonant (due to particles { parallel} streaming and magnetic drift) and non-resonant (due to Larmor radius effects). In Section~\ref{sec:loc_desc_GK} we start from the linearised gyrokinetic equation and transform it into a linear system of equations, introducing the key ordering in the curvature drift and localisation of the mode. In the following section, Section~\ref{eqn:gauss_mode_model}, we focus on constructing the dispersion equation for a model simple Gaussian shaped mode, and detail its derivation and numerical solution. Section~\ref{sec:phys_loc_ITG} then studies the physical elements of the model constructed, including the roles of Landau damping, finite Larmor radius (FLR) stabilisation and the curvature drift resonance. Estimates for important features such as critical temperature gradients are also given. Section~\ref{sec:high_harm} then discusses the behaviour of other modes other than that with the simplest Gaussian structure, to end with some numerical gyrokinetic simulations. Although the model is too simple to be quantitatively correct, we show it serves as a blueprint to interpret the results from the simulations. We use this to make connections to the presence of de-localised (slab) and hopping modes. 

% We then discuss the ITG localisation properties in the context of \textit{stellarator optimisation}, focusing on how modifying the field's geometry (often beyond the commonly studied regimes) affects the instability. 

\section{Localised description of the gyrokinetic equation} \label{sec:loc_desc_GK}

% In the fluid limit, the description of the ITG mode along the field line takes the form of a second order ODE in $\ell$. In a sense, it corresponds to a description of a wavefunction in a quantum well whose shape is determined by the geometry of the field. The question here is whether a similar construction is possible, but one in which kinetic effects concerning the dynamics along the field-line are included. In particular, we would like to retain the Landau damping, which we see as an important stabilising mechanism of the turbulence. 

\subsection{Rewriting the gyrokinetic equation}
The starting point of our model construction is the linear gyrokinetic (GK) equation, which we write as follows for the ions,
\begin{equation}
    iv_\parallel\partial_\ell g + (\omega - \tilde{\omega}_d)g = \frac{q_i}{T_i}F_{0i} J_0(\omega-\tilde{\omega}_\star)\phi. \label{eqn:GK}
\end{equation}
The equation is to be understood as a first order ordinary differential equation in $\ell$ \citep{Bryan}, the distance along a field-line, for the non-adiabatic part of the distribution function, denoted by $g$, which is a perturbation { with} respect to the background ion Maxwellian $F_{0i}$. To write the equation in this one-dimensional form, the ballooning transform has been considered \citep{Bryan,CHT,TCH,antonsen1980kinetic} so that $\mathbf{k}_\perp=k_\alpha\nabla\alpha+k_\psi\nabla\psi$. Here $\alpha$ and $\psi$ are defined to be the straight field line Clebsch variables \citep{d2012flux}, such that $\mathbf{B}=\nabla\psi\times\nabla\alpha$ and the toroidal magnetic flux is $2\pi \psi$. The response of the system is coupled to the electrostatic potential $\phi$, and is modulated by the geometry of the magnetic field. Elements of the geometry are present in $\tilde{\omega}_d$ and $J_0$. The former represents the ion magnetic particle drifts, which in the small $\beta$ limit may be written as $\tilde{\omega}_d=\omega_d(x_\parallel^2+x_\perp^2/2)$, where $\omega_d=\mathbf{v}_D\cdot\mathbf{k}_\perp=v_{Ti}\rho_i\kappa\times\mathbf{B}\cdot\mathbf{k}_\perp/B$, where { $x_\parallel=v_\parallel/v_{Ti}$ and $x_\perp=v_\perp/v_{Ti}$ are velocities normalised to the ion thermal speed $v_{Ti}=\sqrt{2T_i/m_i}$, and $\rho_i=m_i v_T/q_iB$ is the ion Larmor radius, where $T_i$, $q_i$ and $m_i$ are the ion temperature, charge and mass.} For simplicity, we will  focus on the simpler limit $k_\psi=0$. The second element of geometry is included in the finite Larmor radius term $J_0=J_0(x_\perp\sqrt{2b})$, where $J_0$ is the Bessel function of the first kind, and $b=(k_\perp\rho_i)^2/2$. The drive of the instability is  included in the diamagnetic drift, represented here by $\tilde{\omega}_\star=\omega_\star[1+\eta(x^2-3/2)]$, where $\eta=\mathrm{d}\ln T_i/\mathrm{d}\ln n_i$ is the ratio of ion temperature to ion density gradients, and $\omega_\star=(k_\alpha T_i/q_i) (\mathrm{d}\ln n_i/\mathrm{d}\psi)$. 
\par
Because the GK equation is a first order ODE, a solution for $g$ can be written in its most general form using an integrating factor, as originally presented by \cite{connor1980stability}. However, the resulting integral expressions, in their generality, do not always manifest clearly the role played by the different physical elements in the problem. After imposing quasineutrality, the relation between the mode structure and the degree of instability is not clear, as an involved integral eigenvalue problem ensues \citep{romanelli}. To circumvent this complexity, we present here an approach that makes the resonant kinetic problem as close as possible to a second order ordinary differential equation, much in the way that it occurs in the fluid limit \footnote{Some classes of integral eigenvalue equations can be easily related to linear differential problems \citep{tricomi_integral}, but this is not straightforward in our kinetic case.}. 
\par
To that end, we first invoke symmetry arguments to simplify the construction of the solution as much as possible. Given the explicit involvement of $v_\parallel$ in the GK equation, it is convenient to separate $g$ into even and odd parts in $v_\parallel$, namely,
\begin{subequations}
    \begin{gather}
        g^e(\ell,\mathbf{v}) = \frac{1}{2}\left[g(v_\parallel)+g(-v_\parallel)\right], \\
        g^o(\ell,\mathbf{v}) = \frac{1}{2}\left[g(v_\parallel)-g(-v_\parallel)\right].
    \end{gather}
\end{subequations}
With these well-defined parity functions, the original GK equation, which had mixed parity, can be separated into two coupled first order differential equations,
\begin{subequations}
    \begin{align}
        iv_\parallel\partial_\ell g^o+(\omega-\tilde{\omega}_d)g^e &=\frac{q_i}{T_i}F_{0i}J_0(\omega-\tilde{\omega}_\star)\phi, \\
        iv_\parallel\partial_\ell g^e+(\omega-\tilde{\omega}_d)g^o &= 0.
    \end{align}
\end{subequations}
The latter is used to eliminate $g^o(\ell,\mathbf{v})$ from the former, to give an equation that only involves the even part of $g$ in $v_\parallel$,\footnote{ A word of caution should be raised about the presence of the resonant denominator $\omega-\tilde{\omega}_d$. The GK equation Eq.~(\ref{eqn:GK}) can be understood as a Laplace transformed version of the its original time dependent form. It is thus well defined for $\Im\{\omega\}>0$ (the usual Bromwich contour), which avoids the `division by zero'. The extension to the remainder of $\omega$-space is treated in Sec.~\ref{sec:simp_disp_fun}. Retaining the resonance is important, especially as it can generate a finite imaginary part of a marginally stable fluid limit.}
\begin{equation}
    v_\parallel\partial_\ell\left[\frac{v_\parallel}{\omega-\tilde{\omega}_d}\partial_\ell g^e\right]+(\omega-\tilde{\omega}_d)g^e=\frac{q_i}{T_i}F_{0i} J_0(\omega-\tilde{\omega}_\star)\phi. \label{eqn:GK_schrod}
\end{equation}
We must not forget that the linearised GK equation, Eq.~(\ref{eqn:GK}), does not come on its own. First, $g^e$ must satisfy vanishing boundary conditions at $\ell\rightarrow\pm\infty$ for a physically reasonable ballooning solution of the equation \citep{CHT}. Second, it must be complemented by the quasineutrality condition, the condition preventing charge separation from building up in the system. This imposes an additional relation between the velocity-space function $g^e$ and the real space function $\phi$, which is necessary to complete the $\omega$-eigenvalue equation. As the velocity-space average of the odd $g^o$ vanishes due to parity, the quasineutrality condition reduces to,
\begin{equation}
    \int J_0 g^e \mathrm{d}^3\mathbf{v} = \bar{n}(1+\tau)\phi, \label{eqn:QN_schrod}
\end{equation}
where $\tau=T_i/ZT_e$ and $\bar{n}$ is the equilibrium density. We are taking the electron response to be adiabatic here. Equations~(\ref{eqn:GK_schrod}) and (\ref{eqn:QN_schrod}), describing our ion response, will be our starting point. 

\subsection{Approximations: localisation and weak curvature}
So far, the problem defined by Eqs.~(\ref{eqn:GK_schrod}) and (\ref{eqn:QN_schrod}) is as general as the standard form of the linearised GK equation.  To proceed, we introduce a number of simplifying assumptions that will make it tractable while retaining the main physical elements of the problem.
\par
\subsubsection{Simplified geometry}
The first element of simplification in the problem is the geometry along the field line. We restrict all the inhomogeneity in the field to the $\ell$ dependence of the curvature drift $\omega_d(\ell)$. Doing so lets us capture the key aspect of having good and bad curvature regions along the field line. On physical grounds \citep{Terry_And_Horton},  we expect the unstable toroidal mode to have a tendency to localise around bad curvature regions, which are energetically favourable, while being repelled from good curvature ones. 
\par
To introduce a sense of this feature, we describe a single region of bad curvature, taken to be symmetric in $\ell$ and to have a finite extent, $\Lambda$, beyond which the curvature changes sign. In the usual jargon, this length-scale may be deemed the \textit{connection length}, and the region within can be thought as a \textit{bad curvature well}. We model the well as a two-parameter simple symmetric quadratic function in $\ell$, where the parameters represent its depth and width. The magnitude of the bad curvature at the origin is defined to be $-\bar{\omega}_d<0$. The curvature remains \textit{bad}, i.e., negative, in the region $|\ell|<\Lambda$. We then write $\omega_d$ as
\begin{equation}
    \omega_d=\bar{\omega}_d\left[\left(\frac{\ell}{\Lambda}\right)^2-1\right]. \label{eqn:def_wd_model}
\end{equation}
Because $\Lambda$ is the only existing length-scale along the field line, as both $B$ and $k_\perp\rho_i$ are assumed to be constant, we shall use it to normalise lengths and write $\bar{\ell}=\ell/\Lambda$. 
\par
Before moving on, we should reflect on the consequences of our simplified geometry. Choosing the magnetic field magnitude to be flat along the field line eliminates any contribution from trapped ions, which do not exist in our model. The physics associated to the variation of $k_\perp\rho_i$ are also lost, and with it the possible modulation of finite Larmor radius effects. In particular, this approximation erases the effects of \textit{global magnetic shear}, which would have appeared as a secular term in $k_\perp$. The global shear can become important especially in localising significantly extended, slab-like modes, and thus forcing the right behaviour of $g$ at $\ell\rightarrow\pm\infty$. Instead, in our problem, in the limit of large $|\bar{\ell}|$ we have a strongly positive good curvature. Physically, we expect this to play a stabilising role for the typical toroidal ITG that precludes modes from becoming completely delocalised. In that sense, our model of $\omega_d(\ell)$ mimics in part the action of global shear.\footnote{The unbounded behaviour of the curvature drift appears in this context as a rather artificial construct. However, note that in the gyrokinetic equation $\omega_d$ has in fact a secular piece in $\ell$. This piece is in fact proportional to the global magnetic shear, and thus there lies an additional connection between the $\omega_d$ behaviour and the global shear. However, the secularity of $\omega_d$ is not quadratic in $\ell$, but rather linear, as follows from $\mathbf{k}_\perp\sim k_\alpha\nabla\alpha\sim -\iota' k_\alpha\varphi\nabla\psi$.} Unavoidably, this simplification couples the local and global behaviour of the system. All the simplifications considered will render less accurate a quantitative comparison of the analytical results to real fields, but will serve as an important qualitative and semi-quantitative tool. We prove numerically that this particular approximation gives excellent results for the real geometry of the quasisymmetric \citep{nuhren1988,boozer1983transport,rodriguez2020necessary} HSX stellarator \citep{anderson1995}.
\par
Reducing all the field inhomogeneity to $\omega_d$ is a significant formal simplification, making the differential operator $\partial_\ell$ commute with everything in Eq.~(\ref{eqn:GK_schrod}) except: $g$ (we shall drop the superscript $e$ describing parity for simplicity), $\phi$ and $\omega_d$. The assumption of a constant $B$ is of particular importance here. The partial derivative respect to $\ell$ is meant to be taken keeping the velocity space variables $\mu=m_iv_\perp^2/2B$ and $\mathcal{E}=m_iv^2/2$ constant. Only when $B$ is constant is $v_\parallel$ independent of real space.  
\par
With this observation, Eq.~(\ref{eqn:GK_schrod}) becomes, upon commutation of the differential operator
\begin{equation}
    \left(1-\tilde{\omega}_d/\omega\right)^{-1}\frac{v_\parallel^2}{\omega^2}\partial_\ell^2 g+\left(1-\frac{\tilde{\omega}_d}{\omega}\right) g+\frac{v_\parallel^2}{\omega^2}\frac{\partial_\ell\left(\tilde{\omega}_d/\omega\right)}{\left(1-\tilde{\omega}_d/\omega\right)^2}\partial_\ell g=\frac{q_i}{T_i}F_{0i}J_0\left(1-\frac{\tilde{\omega}_\star}{\omega}\right)\phi. \label{eqn:GK_schrod_exp_all}
\end{equation}
We have a second order ODE in which all the explicit spatial dependence can be expressed in terms of polynomials in $\ell$. 

\subsubsection{Weak curvature drift, {strong drive} and strong localisation}
It is clear from the above equation that the drift frequency in this scenario plays a central role in prescribing the behaviour of the instability. The mode will adapt to the geometry described by $\omega_d(\bar{\ell})$, and thus to make the treatment more manageable, it is natural to consider ordering the drift. We introduce the ordering parameter $\delta\sim\bar{\omega}_d/\omega\ll1$. Such a restriction is not completely artificial, and it is particularly appropriate in scenarios in which the turbulence is strongly driven, namely, in cases where the $\omega_\star$ drive (either in its density or temperature gradient form) is much larger than $\omega_d$. We shall be focusing on this strongly driven scenario.
\par
Note however that ordering $\bar{\omega}_d/\omega$ is not enough to simplify Eq.~(\ref{eqn:GK_schrod_exp_all}). Although $\bar{\omega}_d/\omega$ may be small, this is only its value at the bottom of the bad curvature well. Thus, there always exists a sufficiently large value of $\bar{\ell}$ such that $\omega_d/\omega\gtrsim 1$. This appears hopeless for an approximated approach to Eq.~(\ref{eqn:GK_schrod_exp_all}). However, we must not consider $\omega_d$ on its own when doing so, but rather alongside $g$ and $\phi$. If the distribution function and the potential are finite only over some finite length scale, then the effective value of $\bar{\ell}$ should reflect that finite extent. Considering a Gaussian-like envelope of the form $\sim e^{-\lambda\bar{\ell}^2/2}$, so that $g$ and $\phi$ have lengthscales $\sim1/\sqrt{\Re\{\lambda\}}$, { where $\Re\{\lambda\}$ denotes the real part of $\lambda$ which is generally complex}, we limit this problem with a new ordering parameter,
\begin{equation}
    \epsilon=\frac{1}{\Re\{\lambda\}}\frac{\bar{\omega}_d}{\omega}\ll1. \label{eqn:def_epsilon}
\end{equation}
The parameter $\epsilon$ limits the construction that follows to modes that are sufficiently localised. Modes may have some degree of de-localisation, but limited by the ordering parameter $\delta$. We thus shall be managing two simultaneous ordering assumptions, namely $\delta,~\epsilon\ll1$. The precise implications of these orderings and how we proceed with them is detailed in what follows.
\par
First, let us employ these newly introduced approximations to simplify Eq.~(\ref{eqn:GK_schrod_exp_all}). Expanding the equation to, and keeping, order $O(\epsilon,~\delta)$,
\begin{equation}
    2\left(\frac{\omega_tx_\parallel}{\omega}\right)^2\partial_{\bar{\ell}}^2 g+\left(1-\frac{\tilde{\omega}_d}{\omega}(\bar{\ell}^2-1)\right)g+4\tilde{\omega}_d\left(\frac{\omega_tx_\parallel}{\omega}\right)^2\bar{\ell}\partial_{\bar{\ell}}g=\frac{q_i}{T_i}F_{0i}J_0\left(1-\frac{\tilde{\omega}_\star}{\omega}\right)\phi, \label{eqn:GK_schrod_herm}
\end{equation}
where allowing some looseness in the notation, $\tilde{\omega}_d=\bar{\omega}_d(x_\parallel^2+x_\perp^2/2)$ here, and $\omega_t$ is the transit frequency defined as,
\begin{equation}
    \omega_t=\frac{v_{Ti}}{\Lambda\sqrt{2}}. \label{eqn:wt_def}
\end{equation}
As of now, it is not clear what the consistent ordering of $\omega_t/\omega$ must be, but we may deal with that explicitly later. For now, we take it to be order 1, {but consider $\bar{\omega}_d/\omega$ corrections to the first term in Eq.~(\ref{eqn:GK_schrod_herm}) small}.
\par
Localised solutions to a second order ODE like Eq.~(\ref{eqn:GK_schrod_herm}) may be approximated by considering a representation of $g$ and $\phi$ in a Taylor-Gauss basis,
\begin{equation}
    g=\sum_{n=0}^\infty \frac{g_n\bar{\ell}^n}{N(n)} e^{-\lambda\bar{\ell}^2/2}, \label{eqn:g_exp_herm}
\end{equation}
where $N(n)=\sqrt{n!/\Re\{\lambda\}^n}$ is a normalisation factor. The Taylor part of the basis (i.e., the expansion in powers of $\bar{\ell}$) naturally describes the mode near the bottom of the well, while the Gaussian part provides an overall envelope that localises the mode. The latter requires $\Re\{\lambda\}>0$, although it is consistent with having a non-zero imaginary part.\footnote{It could be tempting to use Hermite polynomials instead of unpaired powers of $\bar{\ell}$ as has been done to deal with velocity space in the literature \citep{mandell2018laguerre}. However, the Taylor form is more convenient here with the consideration of the solution in the limit $\bar{\ell}\rightarrow0$ in mind.} The normalisation factor includes powers of $\Re\{\lambda\}$ to account for the scale associated to the monomial $\bar{\ell}^n$. Note that the higher the mode considered, the increasingly hollower the shape of the mode is, providing a characteristic length-scale $\sqrt{n}/\Re\{\lambda\}^{n/2}$. 
\par
Once we have this basis representation, the ordinary differential equation Eq.~(\ref{eqn:GK_schrod_herm}) becomes an infinite set of coupled algebraic linear equation on $\{\phi_n\}$ and $\{g_n\}$. The benefit of the particular basis used is the simplicity of the form that the operations in Eq.~(\ref{eqn:GK_schrod_herm}) take. The product by $\bar{\ell}^2$ { in the second term of Eq.~(\ref{eqn:GK_schrod_herm})} simply upshifts the mode number by 2, making the regularising role of $\epsilon$ manifest. It restricts the coupling of the different $\{g_n\}$ and $\{\phi_n\}$ modes, and thus is critical in achieving a useful truncation of the linear system of equations. Differentiation plays a similar, albeit more involved, coupling role (see Appendix~\ref{app:exp_HG_GK}). 
\par

\subsection{The Taylor-Gauss form of the gyrokinetic equation}
We are in a position now to resolve Eq.~(\ref{eqn:GK_schrod_herm}) in the new basis of Eq.~(\ref{eqn:g_exp_herm}). This may be achieved by substituting the expansion in Eq.~(\ref{eqn:g_exp_herm}), and applying the coupling rules in Eqs.~(\ref{eqn:d2_TayGau}) and (\ref{eqn:l2_TG}). The details of the derivation may be seen in Appendix~\ref{app:exp_HG_GK}. Upon substitution, the resulting equation takes the form
\begin{subequations}
\begin{equation}
    \sum_{n=0}^\infty E_{n}\sqrt{\frac{\Re\{\lambda\}^n}{n!}}\bar{\ell}^ne^{-\lambda\bar{\ell}^2/2}=0,
    \label{eqn:sum_En}
\end{equation}
where, the general expression for the $n$-th mode equation is,
\begin{multline}
    E_n=2\Re\{\lambda\}\left(\frac{\omega_t x_\parallel}{\omega}\right)^2\sqrt{(n+2)(n+1)}g_{n+2} + \\
    +\left[1+\frac{\tilde{\omega}_d}{\omega} -2\lambda(2n+1)\left(\frac{\omega_tx_\parallel}{\omega}\right)^2+4n\frac{\tilde{\omega}_d}{\omega}\left(\frac{\omega_t x_\parallel}{\omega}\right)^2\right]g_n+\\
    +\frac{2}{\Re\{\lambda\}}\left[\lambda^2\left(\frac{\omega_t x_\parallel}{\omega}\right)^2-\frac{\tilde{\omega}_d}{2\omega}-2\lambda\frac{\tilde{\omega}_d}{\omega}\left(\frac{\omega_t x_\parallel}{\omega}\right)^2\right]\sqrt{n(n-1)}g_{n-2}-\\
     -\frac{q_i}{T_i}F_{0i}J_0\left(1-\frac{\tilde{\omega}_\star}{\omega}\right)\phi_n.
    \label{eqn:GK_schrod_herm_n}
\end{multline}
\end{subequations}
For Eq.~(\ref{eqn:GK_schrod_herm}) to be true, Eq.~(\ref{eqn:sum_En}) must hold for all $\bar{\ell}$, meaning that each of the equations $E_{n}$ must vanish separately. The original ODE becomes this way an infinite dimensional system of linear equations, $\{E_n=0\}$, as promised.
\par
Inspection of the structure of this equation shows two important features of the system. First, that even and odd modes in $\bar{\ell}$ are completely decoupled. This is, of course, a direct consequence of the original form of the equation having well defined parity. Unless otherwise stated, we shall consider the even set, of which its most basic form is a Gaussian mode. The results follow similarly for the odd set. The second noteworthy property is that, within each of these subspaces, the system has a tridiagonal structure. That is, the equation couples every mode to the immediately adjacent ones.

\section{Gaussian mode model of kinetic ITG} \label{eqn:gauss_mode_model}
The resolution of the original equation into the Taylor-Gauss basis leaves us with an infinite set of algebraic equations to solve in phase space (i.e., in $\mathbf{v}$ and $\mathbf{\ell}$). It is our goal now to truncate this hierarchy down to obtain a useful model of the problem.
\par
\subsection{Constructing the dispersion function}
Let us start first by understanding what the structure of the set of equations is when we truncate the system at order $N$. That is, when we set all $g_n,~\phi_n=0$ for $n>N$. In that case the finite set of equations we are left with is,
$$\{E_0(0,2),\dots,E_n(n-2,n,n+2),\dots,E_{N}(N-2,N),E_{N+2}(N)\},$$ 
where each element must vanish and the numbers in parenthesis denote the modes (both in $g$ and $\phi$) involved in the equations. This system of equations may be rewritten by rearranging the first $N/2+1$ equations to solve explicitly for $\{g_n:~n\leq N\}$ in terms of $\{\phi_n:~n\leq N\}$ by appropriate linear combinations. This is generally possible, and an explicit construction to order $O(\epsilon^2)$ is provided in Appendix~\ref{app:high_n_considerations} following Eq.~(\ref{eqn:GK_schrod_herm_n}). We may write,
\begin{equation}
\begin{cases}
\begin{aligned}
    g_n(\mathbf{v})=&\sum_{m=0}^{N}\mathbb{D}^{(N)}_{nm}(\mathbf{v})\phi_m & (n\leq N)\\
    g_N(\mathbf{v})=&\bar{\mathbb{D}}_{N}(\mathbf{v})\phi_N,
\end{aligned}
\end{cases} \label{eqn:gn_Dnm_phim}
\end{equation}
where $\mathbb{D}_{nm}$ and $\bar{\mathbb{D}}_{N}$ are known functions of velocity space. The last isolated equation on $g_N$, coming from $E_{N+2}$, is a result of the truncation at a finite $N$, and leads to $g_N$ satisfying two equations simultaneously.  The lack of degrees of freedom to salvage this overdetermination in velocity space is indicative of our failure to satisfy the original equation exactly with a finite number of modes. After all, we are attempting an approximate solution. We must therefore assess what is the error made by relaxing this last constraint. To do so, take $M$ to be the dominant mode in the system, which we shall take to be $O(1)$. As shown in Appendix~\ref{app:high_n_considerations} explicitly, because the mode coupling terms are order $\epsilon$, then we expect the magnitude of the error made by dropping that last equation to be $\sim\epsilon^{(N-M)/2}$. The error made is thus small provided we restrict ourselves to $M\ll N\ll N_\epsilon=1/\epsilon$. The upper bound $N_\epsilon$ is set to preserve the presumed smallness of $\epsilon$ as it is amplified by the mode number of the solution to our system of equations. If $N$ is chosen to be sufficiently large, the matrix elements $\mathbb{D}_{nm}^{(N)}$ should then become largely independent of $N$, and we may drop the $(N)$ superscript. Having a model that is largely independent of the truncation is appropriate, and we shall see that a special choice of $\lambda$ enacts this truncation most snugly.
\par
Our gyrokinetic problem has this way reduced to the main linear system of equations in Eq.~(\ref{eqn:gn_Dnm_phim}). To complete the problem, we apply quasineutrality, Eq.~(\ref{eqn:QN_schrod}). Taking the appropriate velocity moment of $\mathbb{D}_{nm}$, and defining,
\begin{equation}
    \mathcal{D}_{nm}=\frac{1}{\bar{n}}\int J_0 \mathbb{D}_{nm}(\mathbf{v})\mathrm{d}^3\mathbf{v},
\end{equation}
the resulting eigenvalue problem becomes,
\begin{subequations}
\begin{align}
    \sum_{j=0}^N \mathbb{M}_{ij}\phi_j=0, \label{eqn:sys_eq_GK_n} \\
    \mathbb{M}_{ij}=(1+\tau)\delta_{ii}-\mathcal{D}_{ij}.
\end{align}
\end{subequations}
Once we solve this system of equations, we may then construct $g_n$ for $n<N$, acknowledging the order $\epsilon^{N-M}$ error made as described above.
\par
We now illustrate our way forward for the $M=0$ mode (the approach for a general $M$ is given in Appendix~\ref{app:high_n_considerations}). This is to take the Gaussian centred about the bad curvature to be our dominant mode; formally, $\phi_0\sim O(1)$. Following the principle of staying as distant as possible from the truncation point of the system, we may ask when the solution to the problem is consistent with  $\phi_n=0$ for all $n>0$. We call this a \textit{pure} mode, in so much as it is consistent with a `complete' decoupling from higher mode numbers. 
% See some further discussion later and in Appendix~\ref{app:high_n_considerations}. 
\par
To order $\epsilon$, the system describing this `pure' $n=0$ mode reduces to the following two equations,
\begin{subequations}
    \begin{align}
        \mathcal{D}_{20}=0, \\
        1+\tau-\mathcal{D}_{00}=0, \label{eqn:disp_eq_2}
    \end{align}
\end{subequations}
where the expressions for $\mathcal{D}_{00}$ and $\mathcal{D}_{20}$ to order $\epsilon$ are,
\begin{subequations}
    \begin{align}
        \mathcal{D}_{20}&=\frac{2\sqrt{2}}{\Re\{\lambda\}}\frac{q_i}{T_i}\int\mathrm{d}^3\mathbf{v}J_0^2 e^{-v^2/v_{Ti}^2}\left(1-\frac{\tilde{\omega}_\star}{\omega}\right)\left[\lambda^2\left(\frac{\omega_t x_\parallel}{\omega}\right)^2-\frac{\tilde{\omega}_d}{2\omega}\right], \label{eqn:D20_disp_eq} \\
        \mathcal{D}_{00}&=\frac{1}{\bar{n}}\frac{q_i}{T_i}\int\mathrm{d}^3\mathbf{v} J_0^2F_{0i}\left(1-\frac{\tilde{\omega}_\star}{\omega}\right)\frac{1}{1+\frac{\tilde{\omega}_d}{\omega}-2\lambda\left(\frac{\omega_tx_\parallel}{\omega}\right)^2}. \label{eqn:D_00}
    \end{align}
\end{subequations}
Both of these expressions constitute the governing dispersion relation for our mode. It might appear that these make the problem overconstrained, however, we must bear in mind that $\omega$ is not the only unknown here. The localisation $\lambda$ is as well, and it must be chosen alongside the frequency of the mode. This is analogous to what happens in the fluid limit \citep{hahm1988properties,Jim_lambda,plunk2014,zocco2016}. In fact, a closed form for $\lambda$ may be obtained from satisfying Eq.~(\ref{eqn:D20_disp_eq}). For simplicity, considering the small $b$ limit and the leading order form of the expression in $\delta,~\epsilon$, 
\begin{gather}
    \mathcal{D}_{20}\approx \frac{\sqrt{2}}{\Re\{\lambda\}}\left[1-\frac{\omega_{\star}}{\omega}(1+\eta)\right]\left[\lambda^2-\frac{\bar{\omega}_d\omega}{\omega_t^2}\right]\approx0, \nonumber\\
    \therefore ~\lambda=\sqrt{\frac{\omega\bar{\omega}_d}{\omega_t^2}}. \label{eqn:lam_kin_itg}
\end{gather}
For physically meaningful modes, and to restrict ourselves to $\Re\{\lambda\}>0$, we define the square root in Eq.~(\ref{eqn:lam_kin_itg}) with its branch cut along the negative real axis in the complex plane (its conventional definition). The mode envelope $\lambda$ indicates a balance between the parallel streaming, $\omega_t^2$, and the curvature drift $\omega_d$, as is well known to be behind the localisation mechanism in the fluid limit of the ITG \citep{hahm1988properties,plunk2014, zocco2016}. Upon approaching marginal stability, the mode will tend to become increasingly de-localised, with an oscillating mode structure when it co-rotates with ions. 
\par
The basic scaling of $\lambda$ together with the ordering of $\epsilon$ implies that, $\xi=\omega_t/\omega$ follows $\delta^{1/2}\xi\leq\epsilon\ll1$. It is convenient then to take for the ordering arguments $\xi\delta^{1/2}\sim\epsilon$. This is consistent with the assumptions made to reach this point.  
% The sign of $\omega_d$ is positive by construction (see definition in Eq.~(\ref{eqn:def_wd_model})), and thus a typical ITG mode that co-precesses with the diamagnetic drift becomes de-localised near marginality. 
Although such consistency is reassuring, the precise form of $\lambda$ in Eq.~(\ref{eqn:lam_kin_itg}) is a direct consequence of the assumption of a `pure' mode. This purity assumption is however not whimsical, but particularly representative. Not only is it consistent with the ordering, but it also brings an elegant and efficient closure to our system of equations (see a discussion on this in Appendix~\ref{app:high_n_considerations}), as well as granting the correct asymptotic fluid limit of the dispersion equation, as we shall later discuss. The choice of $\lambda$ may thus be interpreted as a `fluid' choice, with all the approximations that this involves.
% It is thus natural to focus on this illustrative model to understand the interaction between localisation, kinetics and ITG behaviour. 
However, as one may check upon lifting the purity assumption, this remains a simplifying choice. In fact, such a relaxation eliminates in our case the $\mathcal{D}_{20}=0$ constrain, leaving $\lambda$ undetermined.\footnote{A straightforward way of seeing this is by taking the determinant of $\mathbb{M}$ to vanish. To order $\epsilon$ this is equal to the product of the principal diagonal. The $M=0$ mode thus simply requires vanishing of Eq.~(\ref{eqn:disp_eq_2}).} Future work may be devoted to improving the model by determining $\lambda$ in some other way, perhaps treating it as a free parameter to optimise to maximise the growth rate of the ITG. However, in this first work we willingly sacrifice a precise quantitative prediction for simplicity. We will illustrate this expected quantitative mismatch comparing the model to some example gyrokinetic simulations. We shall however see that the model remains a highly useful tool to interpret complicated {linear} turbulence spectra. 
\par
With this simple form for $\lambda$ as a function of $\omega$, we  then interpret Eq.~(\ref{eqn:D_00}) as a dispersion relation for  $\omega$,
\begin{equation}
    \mathcal{D}=1+\tau+\frac{\zeta}{\bar{n}}\int J_0^2\left(1-\frac{\tilde{\omega}_\star}{\omega}\right)\frac{f_0}{x_\parallel^2-\zeta\left(1+\frac{\bar{\omega}_d}{2\omega}x_\perp^2\right)}\mathrm{d}^3\mathbf{v}, \label{eqn:disp_fun_kin_itg}
\end{equation}
where,
\begin{subequations}
\begin{equation}
    \zeta=\frac{1}{2}\left[\lambda\left(\frac{\omega_t}{\omega}\right)^2-\frac{\bar{\omega}_d}{2\omega}\right]^{-1}. \label{eqn:zeta_def_or}
\end{equation}
or using the form of $\lambda$ in Eq.~(\ref{eqn:lam_kin_itg}),
\begin{equation}
    \zeta=\frac{\omega^2}{2\lambda(1-\lambda/2)\omega_t^2}.
\end{equation}
\end{subequations}
The parameter $\zeta$ can be interpreted as a measure of the kinetic effects, being important for $|\zeta|\sim1.$  Equation~(\ref{eqn:zeta_def_or}) includes resonant kinetics that can come in either through a Landau-type resonance involving the structure along the field line (the $\lambda$ piece), or the magnetic drift resonance.  The relative importance of these terms will depend on the scale of the mode structure along the field line, i.e., the magnitude of $|\lambda|$

% \begin{subequations}
%     \begin{align}
%         \left[-2\lambda(1-\lambda)\left(\frac{\omega_t x_\parallel}{\omega}\right)^2+1-4\lambda\frac{\tilde{\omega}_d}{\omega}\left(\frac{\omega_t x_\parallel}{\omega}\right)^2\right]g_0 &= f_0J_0\left(1-\frac{\tilde{\omega}_\star}{\omega}\right)\phi_0, \label{eqn:kin_itg_i} \\
%         \left[\lambda^2\left(\frac{\omega_t x_\parallel}{\omega}\right)^2-\frac{\tilde{\omega}_d}{\omega}-2\lambda\frac{\tilde{\omega}_d}{\omega}\left(\frac{\omega_t x_\parallel}{\omega}\right)^2\right] g_0 &= -\frac{\tilde{\omega}_d}{2\omega}f_0J_0\left(1-\frac{\tilde{\omega}_\star}{\omega}\right)\phi_0, \label{eqn:kin_itg_ii}
%     \end{align}\label{eqn:kin_itg_both}
% \end{subequations}

\subsection{Simplifying the dispersion function} \label{sec:simp_disp_fun}
To evaluate the dispersion relation in Eq.~(\ref{eqn:disp_fun_kin_itg}) we must perform the necessary velocity space integrals, which includes resolving a resonant denominator. We shall deal with it by first performing the integral over $x_\parallel$ { (and ignoring the resonance in $x_\perp$, as detailed in Appendix~\ref{app:plasma_disp_function})}. To that end, we need to resolve integrals of the following form
\begin{equation}
    I_{\parallel,\beta}(\zeta)=\frac{1}{\sqrt{\pi}}\int_{-\infty}^\infty x_\parallel^{2\beta}\frac{e^{-x_\parallel^2}}{x_\parallel^2-\zeta}\mathrm{d}x_\parallel.
\end{equation}
Using the difference of squares for the denominator, we rewrite the integral
\begin{equation}
    I_{\parallel,\beta}(\zeta)=\frac{1}{\sqrt[*]{\zeta}}\frac{1}{\sqrt{\pi}}\int_{-\infty}^\infty\frac{x_\parallel^{2\beta}e^{-x_\parallel^2}}{x_\parallel-\sqrt[*]{\zeta}}\mathrm{d}x_\parallel, \label{eqn:I_par_beta}
\end{equation}
which will soon be related to the plasma dispersion function \citep{fried2015plasma}. Here, $\sqrt[*]{\zeta}$ denotes a choice for a branch cut of the function $f(\zeta)=\sqrt{\zeta}$ that maps the portion of the $\omega$-plane $\Im\{\omega\}\rightarrow\infty$ to $\Im\{\sqrt[*]{\zeta}\}>0$. This choice is important for the problem to be consistent with the time-dependent description and the inverse Laplace transform. For large positive growth rates this avoids pole contributions to the Bromwich contour (and thus to the plasma dispersion function), making the inverse Laplace transform well defined. To evaluate the latter it is convenient to deform the Bromwich contour from its original position in the positive imaginary part of the plane downwards. Thus, in addition to the choice regarding the sign of $\Im\{\sqrt[*]{\zeta}\}$, it is convenient to construct a Riemann sheet by placing branch-cuts southwards. 
\par
\begin{figure}
    \centering
    \includegraphics[width=\textwidth]{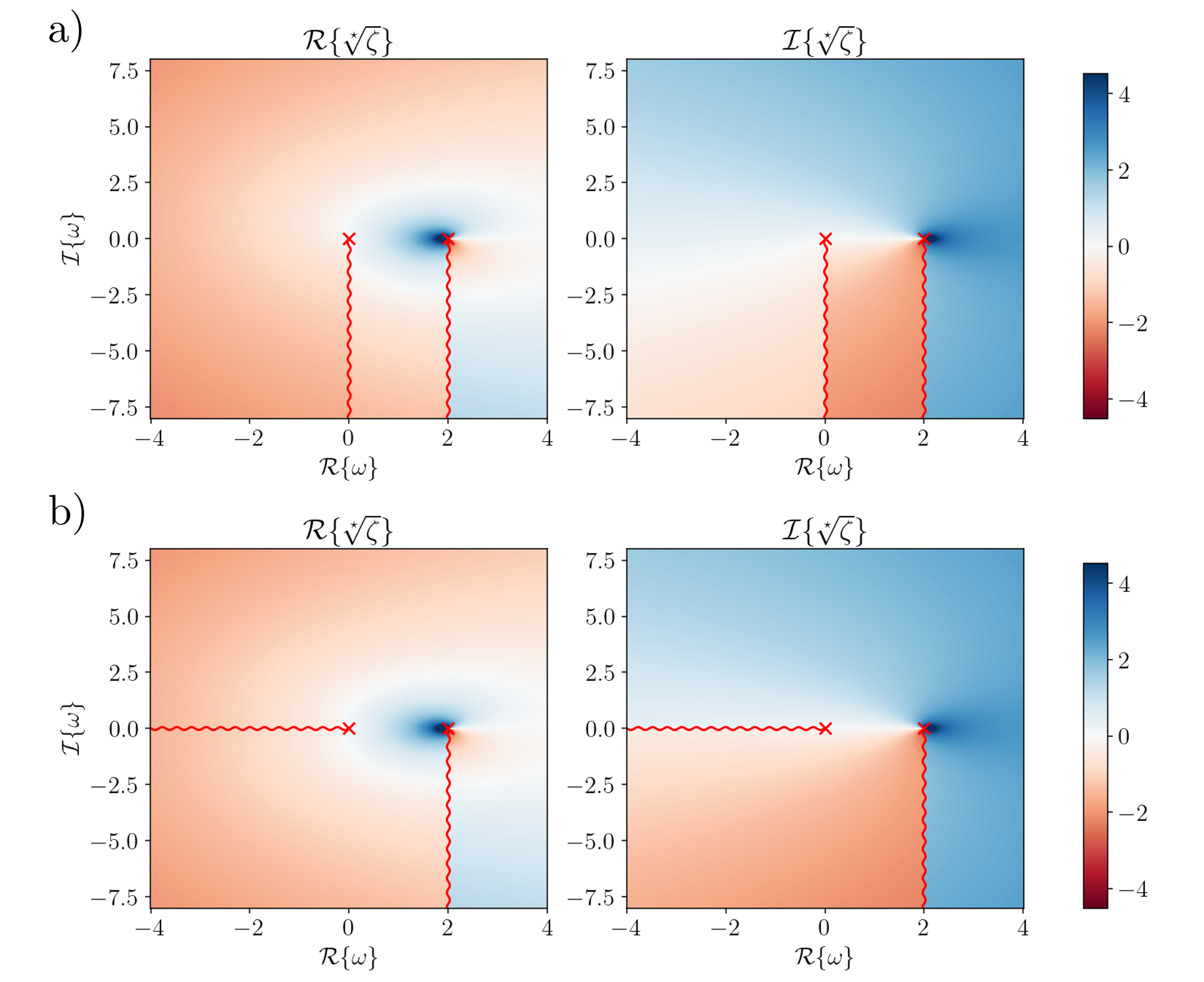}
    \caption{\textbf{Branch cuts for $\sqrt[*]{\zeta}$}. Two different Riemann sheets are shown (a and b) for $\sqrt[*]{\zeta}$ in complex $\omega$-space. The plots on the left and right show the real and imaginary parts of $\sqrt[*]{\zeta}$, respectively. The plots in a) represent the natural Laplace continuation choice for the branch cuts, while b) is the choice that represents localised solutions everywhere in the $\omega$ plane ($\Re\{\lambda\}>0$). The branch cuts are denoted by the red wiggly lines across which the function is discontinuous. The function has an integrable singularity at $\omega=4\omega_t^2/\omega_d$ as indicated in the text. Frequency is normalised to $\omega_t$ and $\omega_t/\omega_d=1/2$ is chosen for illustration, with the colormaps normalised for appropriate visualisation.}
    \label{fig:branch_cut}
\end{figure}
To enforce the above, we choose two branch cuts in $\omega$-space originating from the critical points $\lambda=0$ and $\lambda=1$. The latter is, in addition to a branch point, also a singular point, $\sqrt[*]{\zeta}\sim1/\sqrt{1-\lambda/2}$. The branch points represented will lead to some secular time dependence for damped modes \citep{kuroda1998initial,sugama1999damping} which we shall not explore further in this work. With these branch points localised, the natural branch-cut choice in Figure~\ref{fig:branch_cut}a is problematic, as the branch cut emanating from $\omega=0$ is directly linked to the definition of $\lambda$, Eq.~(\ref{eqn:lam_kin_itg}), and the vertical choice of the cut does not guarantee the physical requirement of localised $\Re\{\lambda\}>0$ modes. To avoid this, one must place the branch cut along the real line, as in Fig.~\ref{fig:branch_cut}b. Although this changes the continuation of the Bromwich contour to the negative $\Im\{\omega\}<0$ part of the plane, it should not affect the description of unstable modes, which is our main concern here.
\par
With either definition in our hands, we proceed and write,
\begin{equation}
    I_{\parallel,\beta}(\zeta)=\frac{(-1)^\beta}{\sqrt[*]{\zeta}} \partial_s^\beta \left[Z\left(\sqrt{s}\sqrt[*]{\zeta}\right)\right]_{s=1},  \label{eqn:I_par_beta_fin}
\end{equation}
where $Z(\cdot)$ is the well known plasma dispersion function \citep{fried2015plasma}, analytical continuation of the integral
\begin{equation}
    Z(a) = \frac{1}{\sqrt{\pi}}\int_{-\infty}^\infty \frac{e^{-x^2}}{x-a}\mathrm{d}x, \label{eqn:plasma_disp_function}
\end{equation}
beyond $\Im\{a\}>0$. The introduction of $s$ as a dummy parameter in Eq.~(\ref{eqn:I_par_beta_fin}) follows from application of the Feynman trick \citep[Ch.~VI]{woods1926advanced} to compute the integral in Eq.~(\ref{eqn:I_par_beta}) in terms of Eq.~(\ref{eqn:plasma_disp_function}).
\par
We then have, after integration over $x_\perp$ (see the details in Appendix~\ref{app:plasma_disp_function}),
\begin{multline}
    \mathcal{D} = 1+\tau + F_0(b)\left[\left(1-\frac{\omega_\star}{\omega}+\frac{3}{2}\frac{\omega_\star^T}{\omega}\right)\sqrt[*]{\zeta}Z(\sqrt[*]{\zeta})-\frac{\omega_\star^T}{\omega}\zeta Z_+\right]-\frac{\omega_\star^T}{\omega}F_2(b)\sqrt[*]{\zeta}Z(\sqrt[*]{\zeta})+\\
    +\frac{\omega_\star\bar{\omega}_d}{4\omega^2}\left\{F_2(b)\left[\eta\left(\frac{3}{2}+\zeta\right)-1+\left(1+2\zeta+\eta\left(2\zeta(\zeta-2)-\frac{3}{2}\right)\right)Z_+\right]+\right.\\
    \left.+F_4(b) \eta \left[(1+2\zeta)Z_+-1\right]\right\}. \label{eqn:disp_kin_ITG_Z}
\end{multline}
where the Larmor radius functions are,
\begin{gather}
    F_0(b)=\Gamma_0, \\
    F_2(b)=(1-b)\Gamma_0+b\Gamma_1, \\
    F_4(b)=2\left[(1-b)^2\Gamma_0+\left(\frac{3}{2}-b\right)b\Gamma_1\right],
\end{gather}
with $\Gamma_n=e^{-b}I_n(b)$, and $I_n$ is the modified Bessel function of the first kind \citep[\S 9.6]{abramowitz1968handbook}. The shorthand notation $Z_+=-Z'/2=1+\sqrt[*]{\zeta}Z(\sqrt[*]{\zeta})$. In this form of the dispersion relation, we have presented all the terms that are, at least explicitly, order 1 or larger, taking $\omega_\star/\omega\sim\omega/\bar{\omega}_d$ in ordering. Additional terms in the expansion could be included (see Tab.~\ref{tab:terms_disp_relation} in Appendix~\ref{app:plasma_disp_function}), although these would stretch the original ordering in $\delta$ and $\epsilon$ and do not include any additional physics. { This dispersion relation describes the behaviour of a linear localised ITG mode responding to a quadratic magnetic curvature with both a good and bad curvature. The dispersion is obtained through ordering  of the localisation and the magnetic drift, which are taken to be strong and weak respectively. In addition, to reach such simple form, we consider what we refer to as a `pure' mode, which simplifies the localisation response of the mode.}
 
\subsection{Computing the dispersion function and finding unstable modes}
Some elements of the dispersion relation are reminiscent of the common local, slab ITG dispersion \citep{kadomtsev1970turbulence} or local, short wavelength ITG \citep{smolyakov2002short}. We extend on these by a more careful consideration of mode localisation.  
\par
\begin{figure}
    \centering
    \includegraphics[width=\textwidth]{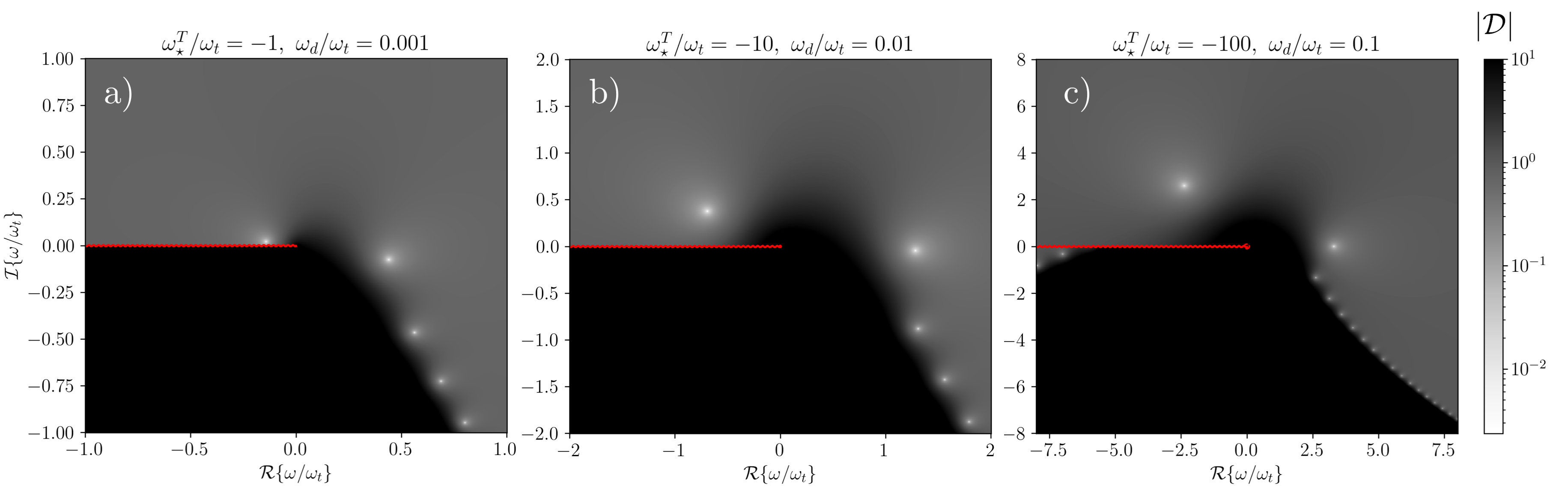}
    \caption{\textbf{Dispersion function $\mathcal{D}$ in $\omega$ space.} The plots show $|\mathcal{D}|$ as a function of complex $\omega$ for different combinations of $\bar{\omega}_d$ and $\omega_\star^T$ (all frequencies normalised to the transit frequency { $\omega_t=v_{Ti}/\Lambda\sqrt{2}$}). The set of three figures may be thus interpreted as the change in the instability due to an increase in $\Lambda$, the width of the bad curvature region, { where the positive $\Im\{\omega\}$ part of the plane denotes instability}. a) $\omega_d/\omega_t=1\times10^{-3}$ and $\omega_\star^T/\omega_t=-1$, b) $\omega_d/\omega_t=1\times10^{-2}$ and $\omega_\star^T/\omega_t=-10$, c) $\omega_d/\omega_t=1\times10^{-1}$ and $\omega_\star^T/\omega_t=-100$. In this case we have chosen $b=0$, $\tau=1$ and $\omega_\star=0$ for simplicity. The red line represents one of the branch cuts; the vertical branch cut is not present in the domain plotted, as the unstable modes live near $\omega=0$ as shown.}
    \label{fig:disp_w_space}
\end{figure}

The information about the linear stability of our localised mode is encoded in the solutions to the dispersion relation $\mathcal{D} = 0$. In general, this constitutes a complicated transcendental equation whose solutions need to be found numerically. The function $\mathcal{D}(\omega)$ can be evaluated in $\omega$-space numerically using the definition of $\sqrt[*]{\zeta}$ above, and using one of the many efficient implementations of the plasma dispersion function (in this case, we use the function \texttt{wofz} \citep{JohnsonFaddeeva} in \texttt{scipy.special}). We accept as valid roots the values of $\omega$ for which $|\mathcal{D}|=0$, which holds true whenever both its real and imaginary parts vanish. In Figure~\ref{fig:disp_w_space} we show some examples of what the dispersion function looks like in $\omega$-space for some particularly simple cases in which no Larmor-radius effects are present.
\par
\begin{figure}
    \centering
    \includegraphics[width=\textwidth]{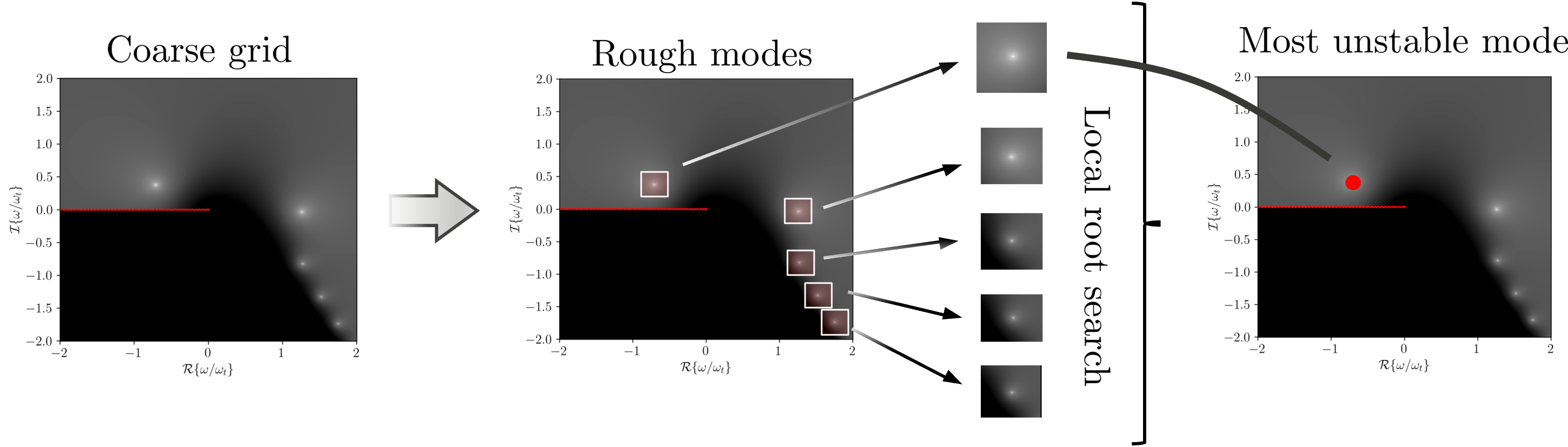}
    \caption{\textbf{Diagram sketching the procedure to find the most unstable mode.} This algorithm is used when numerical roots of $\mathcal{D}$ are required.}
    \label{fig:diagram_disp_roots_num}
\end{figure}

A single unstable mode with $\Re\{\omega\}<0$ is seen in the plots of Fig.~\ref{fig:disp_w_space}. As the drift and diamagnetic frequencies are varied, the mode location evolves in $\omega$-space. In fact, if we interpret the plots (a) to (c) as the change due to increasing $\Lambda$, the width of the bad curvature region along the field line, we see that decreasing $\Lambda$ below a certain threshold appears to lead to the unstable mode eventually vanishing. The evolution of this instability with $\Lambda$ and the other parameters is what we are ultimately interested in, and shall be the main focus of our study in the following section. To automate the root-find of $\mathcal{D}$ and be able to study the various interesting mode dependencies, we implement the following approach: i) we evaluate $|\mathcal{D}|$ in a coarse-grained $\omega$-space roughly bounded by $\omega_\star^T$, ii) we find the regions of lowest $|\mathcal{D}|$ residual, iii) we perform local least-squares minimisation around them, and iv) from the multiple roots found we select the most unstable one (see diagram in Fig.~\ref{fig:diagram_disp_roots_num}). 

% We do so and present the resulting evolution of $\omega$ of this root of $\mathcal{D}$ in Figure~ . The numerical procedure is initialised with evaluating $|\mathcal{D}|$ in a rough complex $\omega$-grid. Once we localise the points with lowest value of $|\mathcal{D}|$, a local non-linear least-squares minimisation is performed. We return the most unstable mode, and ignore any mode with $\mathcal{I}\{\omega\}<0$. An example of 
% the mode evolution with $\Lambda$ is shown in Figure~\ref{fig:itg_kin_Lam_evol}.
% \begin{figure}
%     \centering
%     \includegraphics[width=0.8\textwidth]{figures/disp_Lam_evo_kin_wT_-10.00_wd_-2.00_wn_-0.00_b_0.00.png}
%     \caption{\textbf{Localised, kinetic ITG mode evolution with $\Lambda$.} The plot show the evolution of the dependence of the localised ITG mode on the bad curvature well width $\Lambda$. The plot shows both the real (dashed) and imaginary (solid) parts of $\omega$, normalised to the the transit frequency $\omega_{t0}$ at $\Lambda=\Lambda_0$. The dotted line represents the fluid localised slab ITG limit, while the dash-dotted line corresponds to the fluid toroidal ITG growth rate. A threshold is found in this case at $\Lambda/\Lambda_0\approx0.06$. The curve is computed for $\omega_d/\omega_{t0}=0.01$, $\omega_\star^T/\omega_{t0}=-10$, $\omega_\star/\omega_{t0}=0$, $b=0$ and $\tau=1$.}
%     \label{fig:itg_kin_Lam_evol}
% \end{figure}

\section{Physics and features of the localised kinetic ITG mode} \label{sec:phys_loc_ITG}
In this section we investigate the behaviour of the localised ITG mode, including kinetic effects, by exploring Eq.~(\ref{eqn:disp_kin_ITG_Z}). By inspection,the modes can possibly depend on the following parameters: the diamagnetic drive $\omega_\star$ and $\omega_\star^T$ (density and temperature gradients respectively), the curvature drift magnitude $\bar{\omega}_d$, the transit frequency $\omega_t=v_{Ti}/\Lambda\sqrt{2}$, the poloidal wavenumber $k_\alpha$ and the ratio of electron to ion temperature $\tau$. 
\par
The frequencies and length-scales will be presented normalised to a reference transit frequency, $\omega_{t0}$. That is, the frequency characteristic of the motion of a thermal ion over a distance $\Lambda_0$ along the field line, $\omega_{t0}=v_{Ti}/\Lambda_0\sqrt{2}$. We take this lengthscale $\Lambda_0$ to normalise $\Lambda$ as well. The normalisation of the poloidal wavenumber is somewhat more complicated. Let us recall the definition of $k_\alpha$ from the covariant form of the wavevector $\mathbf{k}_\perp=k_\alpha\nabla\alpha+k_\psi\nabla\psi$. We took for simplicity $k_\psi=0$, so that $\mathbf{k}_\perp=k_\alpha\nabla\alpha$. The parameter $k_\alpha$ is dimensionless, leaving us with the uncomfortable situation of $k_\alpha\rho_i$ having units of length. When we write $k_\alpha\rho_i$ in what follows we adopt the convention of meaning $\overline{k_\alpha\rho_i}=k_\alpha\rho_i|\nabla\alpha|$ (we will often drop the overbar notation, but it should be clear when we do so). The Larmor radius parameter becomes $b=(\overline{k_\alpha\rho_i})^2/2$, which does not exactly match the convention in other works in which the minor radius scale $a$ is chosen to normalise $k_\alpha\rho_i$.
\par
\begin{figure}
    \centering
    \includegraphics[width=\textwidth]{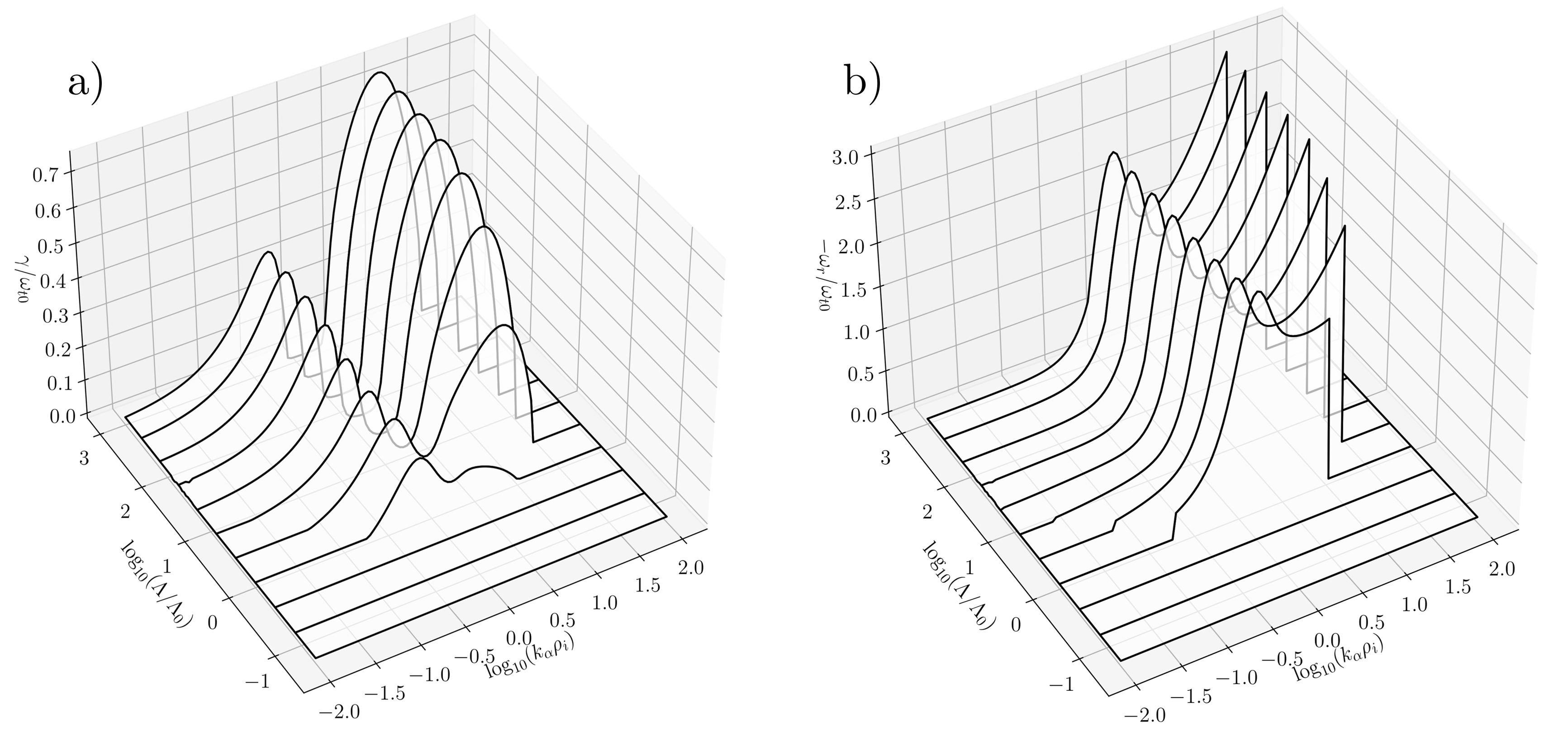}
    \caption{\textbf{Evolution of the {linear} spectrum of unstable ITG with the bad curvature region size, $\Lambda$.} The plots show a) the growth rate and b) negative the real frequency of the most unstable mode as a function of $k_\alpha\rho_i$ and $\Lambda$. The frequencies are normalised to $\omega_{t0}=v_{Ti}/\Lambda_0$, where $\Lambda_0$ is some reference length. We only plot points when the mode satisfies the conditions $\gamma>0$ and $|\omega_d/\omega|<1$. The plots are constructed for the choice $\omega_\star^T/\omega_{t0}=-10$, $\omega_d/\omega_{t0}=0.1$, $\omega_\star/\omega_{t0}=0.0$ and $\tau=1$.}
    \label{fig:3d_Lam_kalpha}
\end{figure}
\begin{figure}
    \centering
    \includegraphics[width=\textwidth]{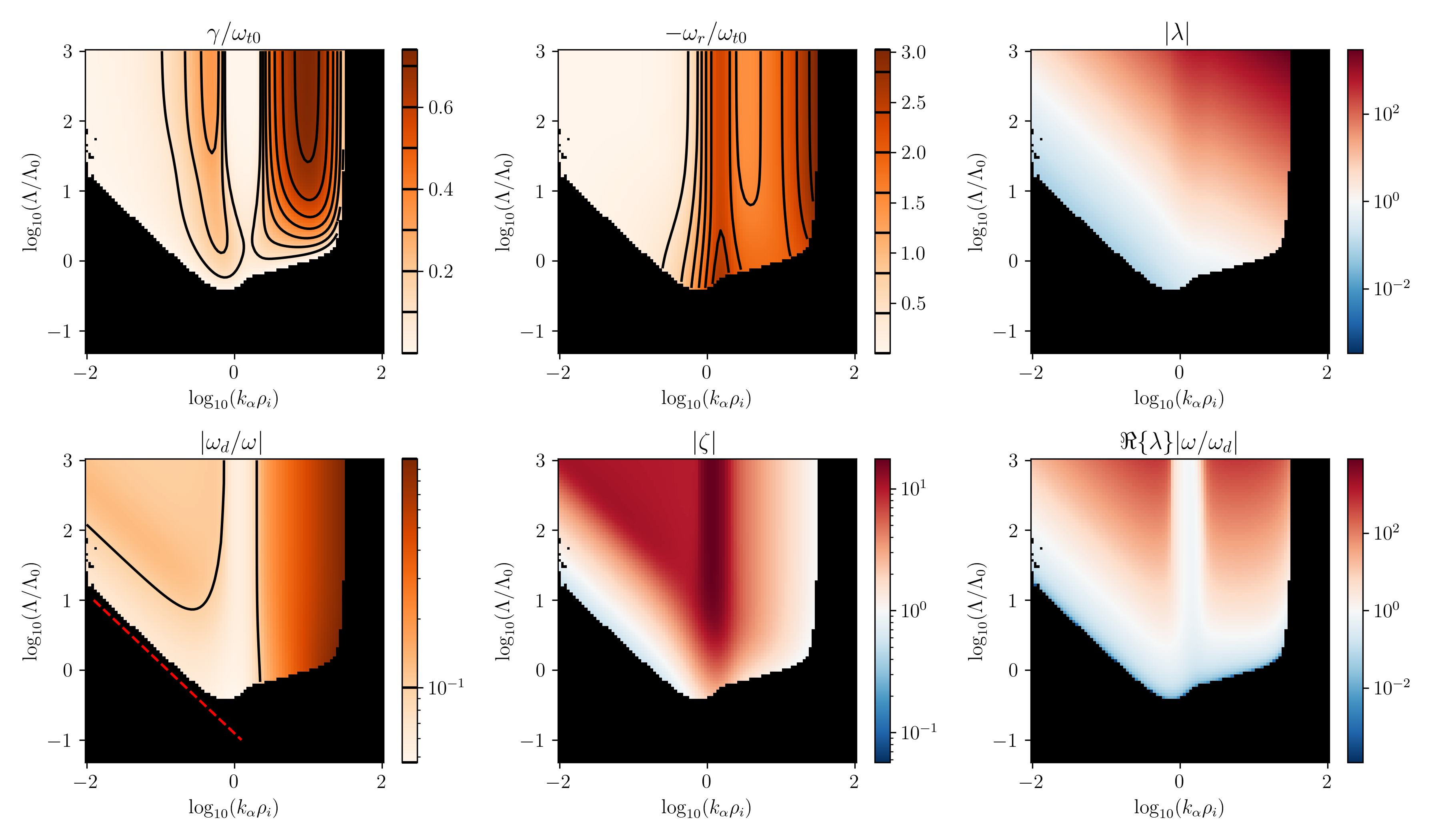}
    \caption{\textbf{Properties of the unstable modes as a function of the bad curvature region size, $\Lambda$, and $k_\alpha\rho_i$ for $\gamma>0$ and $|\omega_d/\omega|<1$.} The plots show, clockwise starting from the top left, the growth rate $\gamma$, the real frequency $-\omega_r$, the Gaussian envelope scale $|\lambda|$, the approximation scale $\Re\{\lambda\}|\omega/\omega_d|$, the kinetic measure $|\zeta|$ and the small scale $|\omega_d/\omega|$. The top left and middle plots can be interpreted as top views of Fig.~\ref{fig:3d_Lam_kalpha}. The red broken line in the bottom left plot is the estimate of the Landau threshold as detailed in Sec.~\ref{sec:landau_thresh}. The blue region in the bottom right plot shows where we expect our localised mode approximation to break down. Here $\omega_\star^T/\omega_{t0}=-10$, $\omega_d/\omega_{t0}=0.1$, $\omega_\star/\omega_{t0}=0.0$ and $\tau=1$.}
    \label{fig:comp_Lam_kalpha}
\end{figure}

We now construct the {linear} spectrum of the localised ITG mode as a function of $\overline{k_\alpha\rho_i}$, as the width of the bad curvature region is changed. Some examples are presented in Figures~\ref{fig:3d_Lam_kalpha} and \ref{fig:comp_Lam_kalpha}. For these cases we have chosen a strongly driven scenario, $\omega_\star^T/\bar{\omega}_d=10^2$, in the presence of no density gradient. 

\subsection{General features and assessment of the approximation}
The {linear} spectra obtained have four distinctive features that are most clear in the limit of a wide bad curvature region.  At small $k_\alpha\rho_i$, as both the diamagnetic drive and the bad curvature diminish (given that they are linear functions of it), so does the instability, Fig.~\ref{fig:3d_Lam_kalpha}a. The mode remains co-rotating with the diamagnetic frequency, Fig.~\ref{fig:3d_Lam_kalpha}b, but eventually, for small enough frequency, it reaches what we may call the \textit{Landau threshold}. This occurs when at some critical $k_\alpha\rho_i$, the frequency matches the parallel transit frequency leading to Landau damping that stabilises the localised mode. As a result, as this threshold is approached and the kinetic aspects of the problem grow ($|\zeta|\rightarrow 1$), the ITG tends to become increasingly stretched over the field line ($|\lambda|$ decreases). As it does so, the ever increasing good curvature of our model (which serves a role similar to that which the magnetic shear would play) dictates its behaviour, as the key approximation $\Re\{\lambda\}|\omega/\omega_d|\sim1/\epsilon\gg1$ fails. This naturally lends to the possibility of the behaviour in this limit being dominated by de-localised, slab-like ITG modes such as \textit{Floquet} modes \citep{zocco2016}. The precise value of $k_\alpha\rho_i$ at which the threshold for the localised mode occurs cannot be expected to be exactly described by the model, but we may use its trends as an informative guide. 
\par
As we increase the poloidal wavenumber, the increase in the diamagnetic and curvature drifts drives the ITG more vigorously. The growth rate increases as the mode becomes more localised; the gains from becoming localised about the bad curvature energy source overcome the effects of increasing the effective $k_\parallel$ that invigorates Landau damping. A larger $k_\alpha\rho_i$ does however also enhance finite-Larmor-radius effects. The stabilisation effects of the geometry can suppress the ITG mode. This leads to the small-$k_\alpha\rho_i$ peak ($k_\alpha\rho_i\lesssim 1$) in the spectrum, Fig.~\ref{fig:3d_Lam_kalpha}a. We may think of this as the \textit{standard ITG peak}, and refer to the threshold (or dip) to its right as \textit{FLR stabilisation threshold}. 
\par
Interestingly, increasing $k_\alpha\rho_i$ does not continue to reinforce the stabilising effect of the field geometry (in our case what one can refer to as flux compression $|\nabla\alpha|\sim1/|\nabla\psi|$). There is some point at which the instability drive beats the FLR effects again, and the mode starts to become more and more unstable. This threshold we refer to as the \textit{FLR weakening threshold}. As the mode keeps growing, it becomes increasingly localised near the bad curvature region, with its characteristic mode frequency rising sharply to settle to a roughly constant value, Fig.~\ref{fig:3d_Lam_kalpha}b. Eventually, the mode reaches a last critical $k_\alpha\rho_i$ threshold. This corresponds to the situation in which the mode resonates with $\omega\sim -\bar{\omega}_d$, Fig.~\ref{fig:comp_Lam_kalpha}. We call this the \textit{$\omega_d$ threshold}. Once again, as the kinetic effects grow and the threshold is approached, the fidelity of the model falters, in this case primarily because $|\bar{\omega}_d/\omega|\sim1$. This leads to a second `hump' in the {linear} spectrum of the ITG instability. Such behaviour has been previously studied in the context of the so-called SWITG instability \citep{hirose2002short,smolyakov2002short,gao2003temperature}.
\par
These features and their location evolve as the width of the bad curvature region changes. They do so in such a way that, as the bad curvature region narrows, the two instability peaks move towards each other, merge and eventually, below a certain critical $\Lambda$, disappear. That is, the localised mode is stabilised by shortening the bad curvature region sufficiently. Unfortunately, and as it occurred near the $k_\alpha\rho_i$ threshold, as $\Lambda$ becomes smaller, the mode becomes increasingly delocalised, and the model presented becomes a poor description. We may expect de-localised modes to gain prominence and persist in this limit.

\subsection{Fluid limit}
Let us start the more quantitative analysis by checking what we obtain in the fluid limit, that is, when all resonances can be neglected. This will prove useful in two ways. First, because it serves as a good check that the dispersion relation reproduces a correct asymptotic limit. Second, because many of the features of our {linear} spectra may be explained in the simplest of terms through the fluid perspective, as we shall see.
\par
With this in mind, let us start by first stating what we mean in this context by the fluid limit: this is the limit of $|\zeta|\rightarrow\infty$, i.e., values far from the kinetic Landau resonance. We therefore need to expand $\mathcal{D}$ in Eq.~(\ref{eqn:disp_kin_ITG_Z}) for large $\zeta$. Since \citep{fried2015plasma}
\begin{equation}
    Z(x)\approx-\frac{1}{x}\left[1+\frac{1}{2x^2}+\frac{3}{4x^4}+\dots\right],
\end{equation}
for large argument, then 
\begin{multline}
    \mathcal{D}\approx 1+\tau-F_0(b)\left[1-\frac{\omega_\star}{\omega}(1-\eta)\right]+\frac{\omega_\star^T}{\omega}F_2(b)-\frac{1}{2\zeta}\left[F_0(b)\left(1-\frac{\omega_\star}{\omega}\right)-\frac{\omega_\star^T}{\omega}F_2(b)\right]+\\
    -\frac{\omega_\star\bar{\omega}_d}{2\omega^2}\left[(1-\eta)F_2(b) + \eta F_4(b)\right].
\end{multline}
where,
\begin{equation}
    \frac{1}{2\zeta}=\left(\frac{\omega_t\sqrt{\lambda}}{\omega}\right)^2-\frac{\bar{\omega}_d}{2\omega}=\frac{\bar{\omega}_d^{1/2}\omega_t}{\omega^{3/2}}-\frac{\bar{\omega}_d}{2\omega}.
\end{equation}
\par
With this explicit form, we may then write the fluid form of the dispersion relation explicitly, 
\begin{multline}
    \mathcal{D}\approx (1+\tau-\Gamma_0)+\frac{\omega_\star}{\omega}\left[F_0(b)(1-\eta)+\eta F_2(b)\right] + \frac{\bar{\omega}_d\omega_\star}{\omega^2}\left[-\Gamma_0+\frac{b}{2}(\Gamma_0-\Gamma_1)-\frac{\eta}{2} F_4(b)\right]+\\
    +\left(\frac{\bar{\omega}_d}{\omega}\right)^{1/2}\frac{\omega_t\omega_\star}{\omega^2}\left[F_0(b)+\eta F_2(b)\right], \label{eqn:gen_b_fluid_eqn}
\end{multline}
where we have dropped terms that are order $\bar{\omega}_d/\omega$. We have an algebraic equation whose roots one can straightforwardly find. Presented in this form of Eq.~(\ref{eqn:gen_b_fluid_eqn}), the dispersion function may { appear} obscure, however it agrees with \cite{connor1980stability} (see Appendix~\ref{app:full_flr_fluid}). This evidences the particular significance of $\lambda$, which provides our construction with the right fluid limit.
\par
It is commonplace to consider the case of a vigorously temperature-gradient driven limit {(namely, $\omega_\star^T/\omega\gg 1$)}, with small  Larmor radius effects, $b\ll1$. We may make a direct comparison in this limit to \cite[Eq.~(19)]{romanelli}\footnote{More precisely, for a fair comparison, one should pick for the ordering $\omega/\omega_\star^T\sim\bar{\omega}_d/\omega\sim(\omega_t/\omega)^2\sim b\sim\delta$. These are the assumptions in \cite{plunk2014}. Note that there is a difference in the sign of $\bar{\omega}_d$ between the two from their respective definition. In fact, this very same limit is achieved even if one ignores the $\omega_\star\bar{\omega}_d$ term in the dispersion function, Eq.~(\ref{eqn:disp_kin_ITG_Z}), although one would not agree with the full Larmor radius expressions in Appendix~\ref{app:full_flr_fluid}.},
\begin{equation}
    \tau\omega^3-b\omega_\star^T\omega^2-\bar{\omega}_d\omega_\star^T\omega+\omega_\star^T\omega_t\sqrt{\omega\bar{\omega}_d}\approx 0. \label{eqn:fluid_eqn}
\end{equation}
The fluid limit of the dispersion function in Eq.~(\ref{eqn:disp_kin_ITG_Z}) is therefore correct. Given this connection, we may extend the common nomenclature distinguishing between the \textit{slab} and \textit{toroidal} ITG modes interpreted as follow \citep{wesson2011tokamaks,zocco2016}. Let the mode in which the streaming terms are ignorable\footnote{Specific orderings were given by \cite{zocco2016}.} be referred to as the \textit{toroidal ITG mode}; and let the \textit{slab ITG mode} correspond to the reverse. Formally, this distinction is dictated by the relative importance of the last two terms in Eq.~(\ref{eqn:fluid_eqn}), which is simply $|\lambda|$. Thus, when the mode has significant structure within the bad curvature well ($|\lambda|>1$), we shall refer to the mode as \textit{toroidal} (note that this mode is not necessarily strongly localised, which would be a statement about $\Re\{\lambda\}$). The \textit{slab} modes will tend to be delocalised and thus care mostly about the larger $|\bar{\ell}|$ good curvature part of the problem.

\subsection{The Landau threshold} \label{sec:landau_thresh}
We discussed qualitatively the presence of a cutoff at long wavelengths $k_\alpha\rho_i\ll 1$ in the {linear} ITG mode spectrum. We argued that this threshold was indicative of a stabilisation of the mode by Landau damping; that is, that the localised mode near the threshold becomes resonant with the parallel streaming frequency. Although very near the threshold the construction of the dispersion relation in Eq.~(\ref{eqn:disp_kin_ITG_Z}) is not formally valid and de-localised modes may exist, we may nevertheless use the model to estimate where this threshold for the localised modes occurs, and how it is affected by the various properties of the field.
\par
We thus need to find a real frequency solution to the equation $\mathcal{D}(\omega)=0$ (i.e., the mode that as in Fig.~\ref{fig:disp_w_space} is just touching the $\omega$-space real axis). Typically, such as in the standard local slab ITG, the real nature of the frequency makes the arguments of the plasma dispersion functions real, and separating the real and imaginary parts of the dispersion relation becomes straightforward, without the requirement of solving any transcendental equation. In the present case, though, the argument of the plasma dispersion function, Eq.~(\ref{eqn:disp_kin_ITG_Z}), involves the square root of $\omega$. As the mode is driven by the temperature gradient, $\omega_\star^T$, $\omega<0$ at the threshold, and thus $Z(\sqrt{\zeta})$ has both mixed real and imaginary parts. This makes solving a trascendental equation unavoidable.
\par
We thus shall proceed by employing a physically motivated approach. As mentioned above, the Landau threshold corresponds to the resonance of the mode with the transit frequency,  $\Re\{\omega\}\sim\omega_t.$ To be more precise, we have  $\zeta\approx1$, and considering the mode to be \textit{slab}-like in the region near the Landau threshold (i.e., rather elongated along the field line compared to the bad curvature region $\Lambda$), we use, { following Eq.~(\ref{eqn:zeta_def_or}),}
\begin{equation}
    \Re\{\omega\}\approx\omega_t\sqrt{2\lambda}=\frac{v_{Ti}}{\Lambda/\sqrt{\lambda}}, \label{eqn:critical_Landau_balance}
\end{equation}
where $\Lambda/\sqrt{\lambda}$ is the characteristic longitudinal scale of the mode structure. The dependence of $\lambda$ on $\Re\{\omega\}$ is known from Eq.~(\ref{eqn:lam_kin_itg}). So what we ultimately need is some notion of the mode frequency, $ \Re\{\omega\}$.
\par
To estimate $ \Re\{\omega\}$, we crudely assume that the kinetic effects mainly affect the stability of the mode, but leave the mode frequency to a large extent unaltered. Assuming a sufficiently large $\Lambda$ and $\omega_\star^T/\bar{\omega}_d$, we use the \textit{slab} branch of the fluid model to estimate $ \Re\{\omega\}$, from Eq.~(\ref{eqn:fluid_eqn}),
\begin{equation}
   \Re\{\omega\}\approx(k_\alpha\rho_i)^{3/5}\hat{\omega}_d^{1/5}\left(\frac{\omega_t\hat{\omega}_\star^T}{\tau}\right)^{2/5}\cos\left(\frac{4\pi}{5}\right), \label{eqn:fluid_localised_slab_itg}
\end{equation}
where the root with a negative real frequency must be chosen given the branch cut choice of the square root. The hat notation is meant to indicate that we have taken the $k_\alpha\rho_i$ dependence of $\bar{\omega}_d$ and $\omega_\star^T$ out explicitly. Although this might appear complicated, and the one fifth powers odd, the expression just found is \textit{precisely} the frequency of a standard slab ITG mode with $k_\parallel=\sqrt{\lambda}/\Lambda$,\footnote{The precise critical threshold that one could estimate using the local slab ITG, using for the mode structure the expression for $\lambda$ in Eq.~(\ref{eqn:lam_kin_itg}), is $k_\alpha\rho_i\approx2\sqrt{2}(1+\tau)(\omega_t/\hat{\omega}_\star^T)\sqrt{\tau\bar{\omega}_d/\omega_\star^T}$. As emphasised in the text, this presents the same basic parameter dependence.}  the characteristic length scale of the mode. The main difference is that in our problem we have explicit field line dependence, while the pure slab ITG does not (other than an oscillating de-localised solution). 
\par
With this expression for the real frequency, we can reconsider Eq.~(\ref{eqn:critical_Landau_balance}), to write
\begin{equation}
    (k_\alpha\rho_i)_\mathrm{Landau}\approx12.5 \tau\frac{\omega_t}{\hat{\omega}_\star^T}\left(\frac{\tau\hat{\omega}_d}{\hat{\omega}_\star^T}\right)^{1/2}. \label{eqn:est_landau_resonance}
\end{equation}
As shown in Fig.~\ref{fig:comp_Lam_kalpha}, this is a fair estimate of the Landau threshold for the parameters considered, and most importantly, it shows the correct $\Lambda$ scaling. The threshold value increases as the bad curvature region becomes smaller, $k_\alpha\rho_i\sim1/\Lambda$. Narrowing down the bad curvature well produces a narrower mode structure (the mode width goes like $\Delta\ell\sim\Lambda^{3/5}$), Landau damping becomes more effective, and thus the threshold increases. If the instability is driven more vigorously (i.e., we increase the temperature gradient drive, $|\omega_\star^T|$), then the parallel dynamics has a harder time stabilising the mode. { The dependence on the temperature gradient of Eq.~(\ref{eqn:est_landau_resonance}) is also in agreement with the numerical solution (see Fig.~\ref{fig:wstarT_kalpha_comp}).} Interestingly, increasing the curvature drift, and with it the magnitude of the bad curvature, does not worsen the threshold, but rather improve it. The reason is that the effect of $\bar{\omega}_d$ in narrowing the mode is, in this limit, stronger than its direct drive of the instability. In this limit in which the mode is rather de-localised, both $\bar{\omega}_d$ and $\Lambda$ should be interpreted to represent more `global' properties of the geometry, as the (global) shear would. 
\par
Of course, near this threshold, the possibilities of other de-localised modes dominating the dynamics increases \citep{zocco_xanthopoulos_doerk_connor_helander_2018,zocco_podavini,podavini2023electrostatic}. In addition, the exact occurrence of the threshold will depend on the precise form of $\lambda$, which we have already acknowledged the current model to only approximate. Thus the particular scaling of the critical $k_\alpha\rho_i$ and its value may change, but the main physical interpretation of its origin and dependence should remain. 

\subsection{The FLR stabilisation}
The next noteworthy feature in the $k_\alpha\rho_i$ spectrum is the stabilisation of the ITG mode as the finite Larmor radius becomes relevant. The result is the appearance of a characteristic peaked {linear} spectrum, with typical values of $k_\alpha\rho_i\lesssim 1$. This feature is not resonant-kinetic, and its presence in Figure~\ref{fig:comp_Lam_kalpha} for large $\Lambda$ indicates it forms part of the fluid description of the instability. 
\par
In the fluid picture, the FLR stabilisation feature of the {linear} ITG spectrum results from the competition between the increased drive of the toroidal ITG and the increased stabilising efficiency of finite Larmor radius effects with increase poloidal wavenumber. Ignoring the streaming term in Eq.~(\ref{eqn:fluid_eqn}), the threshold can be shown to occur when,
\begin{equation}
    \left(\frac{\omega_\star^T b}{2\tau}\right)^2+\frac{\omega_\star^T\bar{\omega}_d}{\tau}\approx0.
\end{equation}
The critical poloidal wavenumber is then,
\begin{equation}
    (k_\alpha\rho_i)_\mathrm{FLR}\approx2\left(\frac{\tau\bar{\omega}_d}{\omega_\star^T}\right)^{1/4}. \label{eqn:crit_FLR}
\end{equation}
From the physical picture above, we expect this stabilisation threshold to have a value close to $k_\alpha\rho_i\sim1$. A sign of this is the weak dependence on both the drift and the temperature gradient. In addition, this mechanism is not directly linked to the mode structure, and hence independent of $\Lambda$. This latter resilience is apparent in Fig.~\ref{fig:comp_Lam_kalpha}. Only a small correction term may be observed within the fluid description of the mode by treating perturbatively the streaming contribution in Eq.~(\ref{eqn:fluid_eqn}), $\delta(k_\alpha \rho_i)_\mathrm{FLR}\approx\tau\omega_t/2|\omega_\star^T|$.
\par
This resiliency of the FLR stabilisation threshold clashes with the rather fundamental dependence of the Landau threshold on $\Lambda$. As a result, we expect the space in $k_\alpha\rho_i$ available to the localised ITG mode to narrow down as the bad curvature region is squeezed. Eventually, the first peak in the {linear} spectrum would be eliminated, as in Fig.~\ref{fig:3d_Lam_kalpha}a, and thus at long wavelengths only extended ITG modes could remain present in the system. From the above, we may estimate the critical width of the bad curvature region, $\Lambda$, at which this occurs. Extrapolating the behaviour of the threshold and the FLR stabilisation, Eqs.~(\ref{eqn:est_landau_resonance}) and (\ref{eqn:crit_FLR}), the balance $(k_\alpha\rho_i)_\mathrm{Landau}\sim(k_\alpha\rho_i)_\mathrm{FLR}$ yields 
\begin{equation}
    \Lambda_\mathrm{crit}\sim6\tau\frac{v_{Ti}}{\hat{\omega}_\star^T}\left(\frac{\tau\hat{\omega}_d}{\hat{\omega}_\star^T}\right)^{1/4}.
\end{equation}
This critical $\Lambda$ could also be rewritten in terms of a critical temperature gradient $(\omega_\star^T)_\mathrm{crit}\sim 4\tau(\omega_t^4\bar{\omega}_d)^{1/5}$. Given the preeminence of the Landau damping physics, the critical temperature gradient threshold below which only extended modes are left at long wavelengths is particularly sensitive to the width of the bad curvature region. This emphasis in controlling the longitudinal spread of the mode aligns with some previous work on critical thresholds \citep{jenko2001critical,roberg2022coarse}.

\subsection{The FLR weakening}
The above considerations of the kinetic threshold and FLR stabilisation explain the presence of a peak in the {linear} spectrum of the mode in $k_\alpha\rho_i$. However, as is clear from Figure~\ref{fig:3d_Lam_kalpha}, the ITG starts growing unstable once again at larger values of the poloidal wavenumber. This might appear surprising at first, because it is natural to think of FLR effects to increase monotonically with $k_\alpha$, and thus after the FLR threshold, its stabilising effect to continue to exceed the increase in the turbulent drive. However, this picture is not correct.
\par
The finite Larmor radius effects become weakened at large $k_\alpha\rho_i$, when the relevant perpendicular scale starts to become significantly smaller than the Larmor radius. Larmor radius effects become inefficient, as they were for small $k_\alpha\rho_i$. Formally, one can ascribe this weakening of the FLR effects to the large argument behaviour of the Bessel functions introduced in the gyrokinetic equation by the gyroaveraging. The small $b$ expansion of the $F_n(b)$ functions fails in this limit, and thus so does the fluid equation written in its typical form of Eq.~(\ref{eqn:fluid_eqn}). Retaining the appropriate behaviour leads to a critical \textit{FLR weakening threshold} beyond which the mode grows back up\footnote{At this scale, the ion dynamics can more efficiently interact with electrons, which in principle should be retained kinetically.  We will not explore these aspects in this work.}. This type of ITG activity is referred to in the literature as short-wavelength ITG (SWITG)  \citep{hirose2002short,smolyakov2002short,gao2003temperature}.
\par
We shall then consider the large $b$ limit of the fluid equation, Eq.~(\ref{eqn:gen_b_fluid_eqn}) using the large argument asymptotics of the Bessel functions \citep[Sec.~9.7]{abramowitz1968handbook}, 
\begin{equation}
    \sqrt{2\pi b}(1+\tau)+\frac{\omega_\star}{\omega}\left(1-\frac{\eta}{2}\right)-\frac{3}{4}\frac{\omega_\star\bar{\omega}_d}{\omega^2}\left(1+\frac{\eta}{2}\right)+\left(\frac{\bar{\omega}_d}{\omega}\right)^{1/2}\frac{\omega_t\omega_\star}{\omega^2}\left(1+\frac{\eta}{2}\right)\approx0. \label{eqn:fluid_eqn_large_b}
\end{equation}
In the limit of $|\lambda|$ being large (which is indeed the behaviour at large $k_\alpha$, see Figure~\ref{fig:comp_Lam_kalpha}), we expect the streaming term in Eq.~(\ref{eqn:fluid_eqn_large_b}) to be small, and thus we are left with, once again, a quadratic in $\omega$. The threshold then occurs when its discriminant vanishes, which gives
\begin{equation}
    (k_\alpha\rho_i)_\mathrm{weak}\approx\frac{1}{3\sqrt{\pi}(1+\tau)}\frac{\omega_\star}{\bar{\omega}_d}\frac{(1-\eta/2)^2}{1+\eta/2}.
\end{equation}
A more refined version of this threshold keeping higher orders in $1/\zeta$ yields for a flat density $(k_\alpha\rho_i)_\mathrm{weak}\approx(0.642-0.032\omega_\star^T/\bar{\omega}_d)/(1+\tau)$. The exact numerical value here is however not important, especially given the ordering in $\delta$ and $\epsilon$ considered. The key is its dependence on the relative magnitude of the diamagnetic and curvature drift, which is quite strong compared to the FLR threshold $(k_\alpha\rho_i)_\mathrm{FLR}\sim(\bar{\omega}_d/\omega_\star^T)^{1/4}$, Eq.~(\ref{eqn:crit_FLR}). Then, as the curvature of the field is increased or the driving temperature gradient reduced, the weakening FLR threshold will approach the FLR stabilisation threshold. Through a simple balance between $(k_\alpha\rho_i)_\mathrm{FLR}\sim(k_\alpha\rho_i)_\mathrm{Weak}$, we expect to find the two regions merging  (see Fig.~\ref{fig:3d_Lam_kalpha}) when $|\omega_\star^T|/\bar{\omega}_d\sim27(1+\tau)^{4/5}\tau^{1/5}$. This is a rather large value of $\omega_\star^T$, meaning that having two distinct peaks in the {linear} spectrum requires of a strongly driven regime (commpared to the drift). 
\begin{figure}
    \centering
    \includegraphics[width=\textwidth]{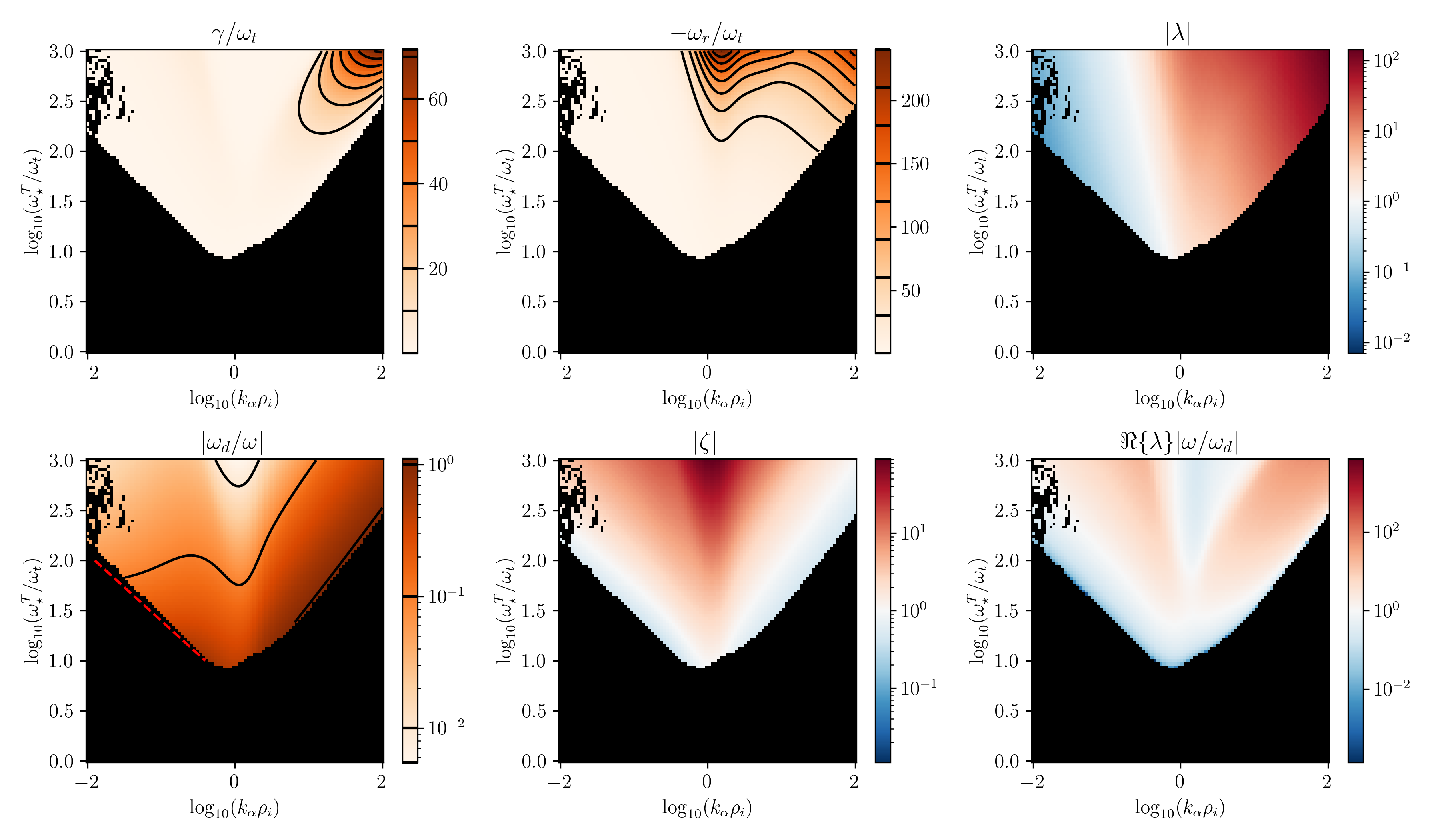}
    \caption{\textbf{Properties of the unstable modes as a function of the temperature-gradient driven diamagnetic frequency, $\omega_\star^T$, and $k_\alpha\rho_i$.} The plots show, clockwise starting from the top left, the growth rate $\gamma$, the real frequency $-\omega_r$, the Gaussian envelope scale $|\lambda|$, the approximation scale $\Re\{\lambda\}|\omega/\bar{\omega}_d|$, the kinetic measure $|\zeta|$ and the small scale $|\bar{\omega}_d/\omega|$. The red broken line in the bottom left plot is the estimate of the Landau threshold as detailed in Sec.~\ref{sec:landau_thresh}. The blue region in the bottom right plot shows where we expect our localised mode approximation to break down. This means that the precise instability threshold in $\Lambda$ cannot be fully trusted. We only plot points when the mode satisfies the conditions $\gamma>0$ and $|\bar{\omega}_d/\omega|<1$. The plots are constructed for the choice $\bar{\omega}_d/\omega_{t}=1.0$, $\omega_\star/\omega_{t0}=0.0$ and $\tau=1$.}
    \label{fig:wstarT_kalpha_comp}
\end{figure}
In Figure~\ref{fig:wstarT_kalpha_comp} we illustrate some of these {linear} spectra dynamics as a a function of a changing temperature gradient. 
% It shows that as the two instability maxima are forced together, the instability critically acquires a more toroidal-like behaviour. This brings the mode towards breaking the $\bar{\omega}_d/\omega$ approximation. It is thus expected that as the critical threshold is approached the behaviour of the mode changes. A different ordering and geometry model would be then necessary. 

\subsection{The $\omega_d$ threshold}
At even larger $k_\alpha\rho_i$ we clearly have another stabilisation effect that leads to the appearance of an instability threshold. The responsible mechanism is not captured in the fluid picture, and is kinetic in nature, as the value of $|\zeta|$ in Figure~\ref{fig:comp_Lam_kalpha} suggests. It could be tempting from our previous discussion on the Landau threshold to suggest that a similar Landau damping mechanism is present here as well. However, as the threshold is approached, the ITG mode does not seem to exhibit a clear tendency to relax its longitudinal structure. In fact, as $|\lambda|\gg1$ the mode retains a fine structure. What is then the mechanism in action? To answer the question it suffices to look back at the definition of $\zeta$ in Eq.~(\ref{eqn:zeta_def_or}), which in this limit gives $\zeta\approx-\omega/\bar{\omega}_d$. Thus, kinetic effects at large $k_\alpha\rho_i$ are dominated by the resonance of the mode rotation with the bad curvature of the field. Note that this resonance has been retained even in our expansion in $\bar{\omega}_d/\omega\sim\delta\ll1$, as can be recognised by inspection of our kinetic equation Eq.~(\ref{eqn:D_00}). 
\par
With the dominant mechanism identified, we may estimate at what wavenumbers the mode frequency balances the drift frequency. The large $k_\alpha\rho_i$ limit of the fluid equation Eq.~(\ref{eqn:fluid_eqn_large_b}) yields a roughly constant real frequency for the mode, which will eventually be matched by the magnetic drift, which grows with the wavenumber $\bar{\omega}_d\propto k_\alpha\rho_i$. Using the higher order cubic version of Eq.~(\ref{eqn:fluid_eqn_large_b}), the balance $\Re\{\omega\}\sim-\bar{\omega}_d$ gives 
\begin{equation}
    (k_\alpha\rho_i)_{\bar{\omega}_d}\approx\frac{1}{2}\frac{1}{1+\tau}\left|\frac{\omega_\star^T}{\bar{\omega}_d}\right|.
    \label{eqn:crit_wd}
\end{equation}
The multiplicative factors in front depend on the details of the approximation of the model, but what is key is the dependence of the threshold on the ratio $\omega_\star^T/\bar{\omega}_d$ (which correctly describes the behaviour in Fig.~\ref{fig:wstarT_kalpha_comp}). The behaviour of this threshold suggests a narrowing of the {linear} spectrum that shall reach a critical narrow range $(k_\alpha\rho_i)^\mathrm{crit}$ when
\begin{equation}
\left.\frac{\omega_\star^T}{\bar{\omega}_d}\right|_\mathrm{crit}\approx2(k_{\alpha}\rho_{i})^\mathrm{crit}\left(1+\tau\right),
\end{equation}
or in tokamak notation $R/L_T$, which is compatible with \cite{romanelli} and \cite{GuoRomanelli} results, with a visual estimate obtained from Fig. (2) of \cite{biglari1989toroidal}, with the Jenko-Dorland-Hammett formula \citep{jenko2001critical}, and other critical gradient estimates \citep{roberg2022coarse}.
\par
In this consideration there appears not to be any direct involvement of the parallel streaming dynamics. However, as the curvature well is narrowed the two kinetic elements in the problem become mixed. Signs of this behaviour are seen in the apparent decorrelation between the $\omega/\bar{\omega}_d$ ratio and the threshold in Figure~\ref{fig:comp_Lam_kalpha} when $\Lambda$ starts having an effect on the mode. In fact, in the above description of a \textit{critical} temperature threshold, we considered some reference critical value of $(k_\alpha\rho_i)_\mathrm{crit}$. Following Figure~\ref{fig:wstarT_kalpha_comp} though, one can expect at some point the $\bar{\omega}_d$ threshold to become close to the Landau threshold, and not just an arbitrarily chosen wavenumber. When this occurs, we may say that there will be no more localised ITG modes. In this case the threshold would scale as,
\begin{equation}
    (\omega_\star^T)_\mathrm{crit}\sim(\bar{\omega}_d^3\omega_t^2)^{1/5}.
\end{equation}
Thus, the threshold changes because we can affect it not only by making the resonance with $\bar{\omega}_d$ appear at longer wavelengths, but also by amplifying the effects involved in the Landau threshold. As a result, increasing the bad curvature or the parallel scale both will have a positive effect on reducing the ITG, although extended modes may persist. The particular scaling obtained by balancing the two physics ingredients goes beyond \cite{biglari1989toroidal} and $R/L_T \sim O(1)$, where $R$ is the major radius. In our case, the balance yields $R/L_T\sim 1/q^{0.6}$, involving the safety factor $q$, in relation to the connection length. The importance of the parallel physics in determining the critical gradients has been recognised by many authors \citep{hahm1988properties,jenko2001critical,roberg2022coarse}, and here we see is involved quite explicitly. Often parallel dynamics are associated to the global shear, which our model does not explicitly treat. However, as discussed with the Landau threshold, the dependence on $\omega_t$ can be related to the role played by global shear when the modes become rather delocalised. The exact form of the scaling with $q$ will depend on the exact behaviour of the mode that is becoming de-localised and thus should be taken with a grain of salt (especially given the weak spot of the model determining $\lambda$). We shall not forget that although increasing the Landau threshold helps in this endeavour of erasing localised modes, one could of course leave behind de-localised modes that could also be deleterious \citep{zocco_xanthopoulos_doerk_connor_helander_2018,zocco_podavini}.

\section{The role of higher harmonics} \label{sec:high_harm}
The kinetic considerations above have focused on the behaviour of a localised mode whose shape is in its simplest form described by a Gaussian envelope. That is, by the `pure' 0-th order in our Taylor-Gauss expansion. Shapes of the modes are seldom so simple, and in the way that we expect multiple modes as solutions to a Schr\"{o}dinger equation (and in fact also the fluid equation \citep{hahm1988properties}), we may also expect to find possible solutions to our problem in which the mode has larger $n$.
\par
A dominant `pure' $M$ mode can be described in a way rather analogous to that considered above for $n=0$. It may be shown, see Appendix~\ref{app:high_n_considerations}, that the resulting equations are identical to the $n=0$ case with the only modification,
\begin{equation}
    \zeta_{n=M}=\frac{1}{2}\left[\lambda(2M+1)\left(\frac{\omega_t}{\omega}\right)^2-\frac{\bar{\omega}_d}{2\omega}\right]^{-1}. \label{eqn:zeta_n_1}
\end{equation}
The only change is the different contribution of the streaming term, formally equivalent to an increased $\omega_t\rightarrow (2M+1)\omega_t$. How can such a scaling be physically interpreted? Let us picture the change to the mode structure as one increases mode number. The mode looks like a pair of peaks pushed against the Gaussian envelope. The larger $M$, the harder they are squeezed, which leads to the width of the peaks to roughly go like $\Delta\bar{\ell}\sim1/\sqrt{M}$, as can be explicitly shown by computing the standard deviation. This linear scaling is what leads to the form in Eq.~(\ref{eqn:zeta_n_1}).
\par
\begin{figure}
    \centering
    \includegraphics[width=0.7\textwidth]{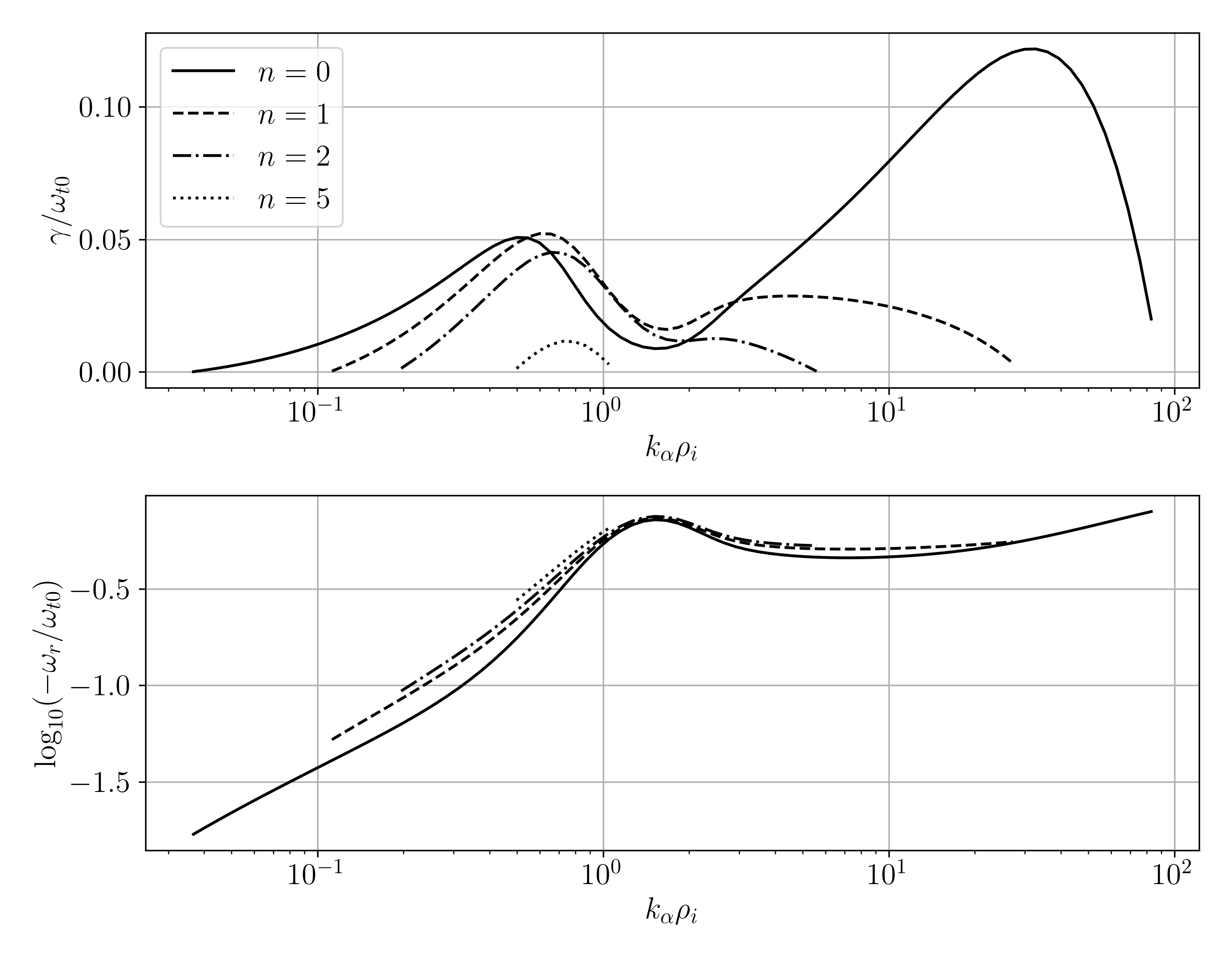}
    \caption{\textbf{Growth and frequency of ITG mode for different structure.} Plots showing the growth rate and real frequency of the ITG mode for the mode numbers $n=0,~1,~2$ using the approximate generalisation of $\zeta$. The plots are computed using $\Lambda/\Lambda_0=10$, $\bar{\omega}_d/\omega_t=0.1$, $\omega_\star^T/\omega_t=-30$, $\omega_\star=0$ and $\tau=1$.}
    \label{fig:enter-label}
\end{figure}

The main effect of $M$ is thus to change the parallel structure of the mode, and thus any of the phenomena that is directly linked to this feature of the instability. The Landau threshold will be most readily affected, but also the particulars of critical thresholds where $\zeta\sim1$ and $|\lambda|$ is not exceedingly large. With the effective $\omega_t$ scaling, we may write $(k_\alpha\rho_i)_\mathrm{Landau}\sim 2M+1$, showing that Landau damping becomes more prominent for the higher modes. This means that we expect to see the lowest modes excited first, as well as those to be most sensitive to stabilisation through squeezing of the connection length, $\Lambda_\mathrm{crit}\sim 2M+1$. The change in the $\omega_d$ threshold with $M$ may appear shocking, given that this we argued is rather insensitive to parallel scales. However, as $M$ is increased, the $\omega_d$ threshold mixes with the Landau threshold, as follows directly from $\zeta$, Eq.~(\ref{eqn:zeta_def_or}). Similar behaviour to what we find here, Fig.~\ref{fig:enter-label}, may be interpreted from the work of \cite{gao2005short}. Thus we expect close to marginality the first localised mode to go unstable to be the $n=0$ mode. 
\par
To study the changes on the magnitude of the growth rate, we write the $M$-th mode generalisation of Eq.~(\ref{eqn:fluid_eqn}), which can be shown to  match other treatments of the fluid equation \citep{hahm1988properties,plunk2014}. In this fluid case, it is clear that increasing mode number enhances the growth rate of our ITG, which is increasingly of a more slab character, $\gamma_\mathrm{slab}\sim(2M+1)^{2/5}$. The fluid picture is unbound! Only through the regularising role played by the kinetic effects is the hierarchy regulated. There is a competition then between larger modes tending towards larger growth rates (in the fluid limit), but also becoming more effectively stabilised by Landau damping. The results of such competition are presented for some example parameters in Figure~\ref{fig:enter-label}. 
\par
These modifications that occur in the model provide us with a flavour of the kind of changes that one would expect when our `pure' mode assumption is relaxed and the true value of $\lambda$ is different.  While the main qualitative features shall remain, we expect quantitative differences to exist. Overall scalings can be understood as in Figure~\ref{fig:enter-label}, but other more sophisticated dependence of $\lambda$ on $k_\alpha\rho_i$ could also lead to additional features in the {linear} spectrum. The model should help us distinguish between these as well.

\section{Qualitative comparison to simulations}
\begin{figure}
    \centering
    \includegraphics[width = \textwidth]{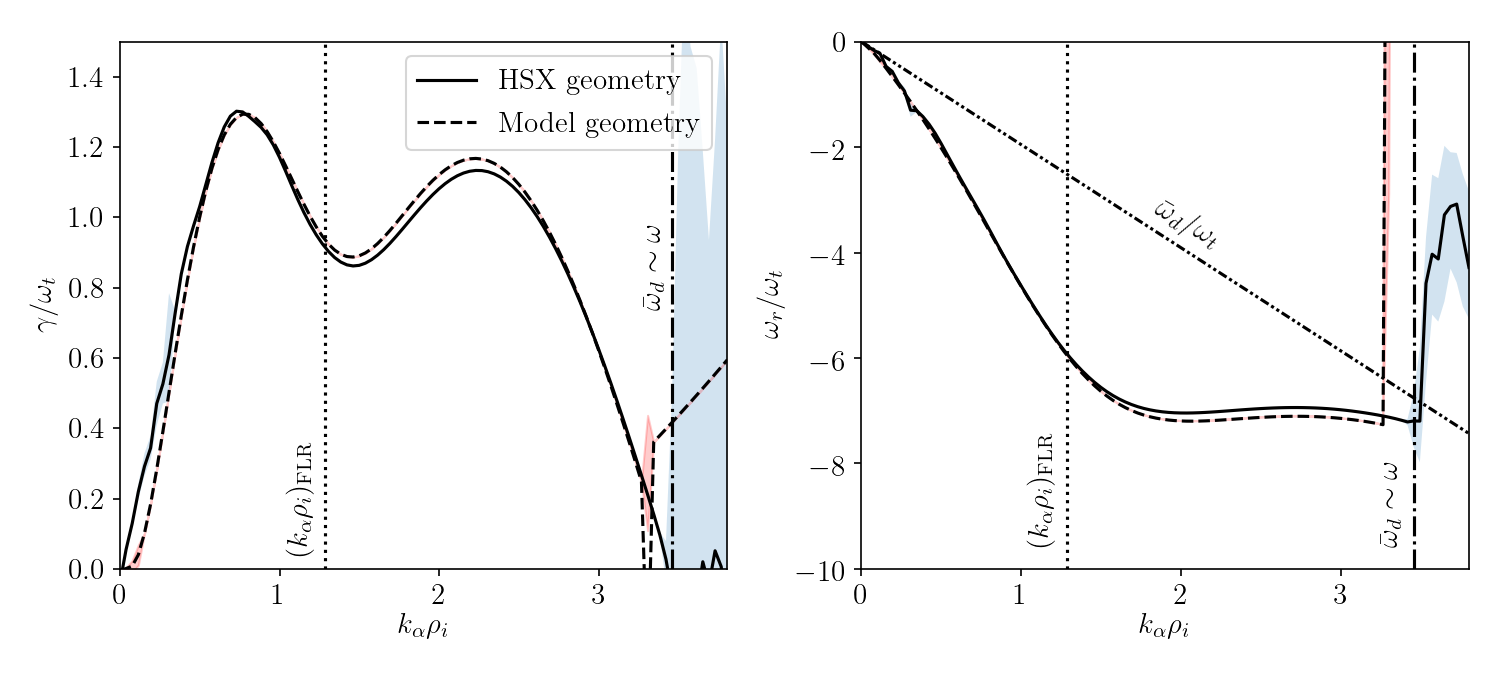}
    \caption{\textbf{{Linear} mode spectrum for HSX gyrokinetic simulations.} The growth rates (left) and real frequencies (right) are shown for linear HSX gyrokinetic simulations with $a/L_T=2.5$ and $a/L_n=0$, and adiabatic electrons. The dashed line corresponds to the simulations performed with { the same gradients but} a modified geometry in which the only spatial dependence in the problem is $\bar{\omega}_d$, and this is modelled as a truncated quadratic well. The coloured shade show the variation in the later half of the simulation of the mode frequency, giving a sense of trust of the spectra (blue and red for the HSX and model geometries respectively). The vertical lines correspond to the predicted FLR stabilisation threshold and the $\omega_d$ stabilising resonance.}
    \label{fig:stella_simulations}
\end{figure}

In this Section we consider as a way of example the {linear} spectrum of ITG modes in a realistic but nonetheless simple stellarator geometry. We use this as way of illustration of how the lessons learnt from our model can be applied in practice to understand behaviour in more complex situations, but also its limitations. The example presented is the {linear} ITG mode spectrum along a flux tube of the HSX \citep{anderson1995} stellarator, a quasisymmetric stellarator \citep{nuhren1988,boozer1983transport,rodriguez2020necessary} with helical symmetry. The latter means that $|\mathbf{B}|$ has a direction of symmetry to prevent fast loss of trapped particles, which implies a particularly simple, quasi-periodic curvature along field-lines. { This simplicity, together with a reduced global magnetic shear, makes this example suitable for the comparison.} The electrostatic mode in the gyrokinetic simulations is driven by an ion temperature gradient $a/L_T=2.5$, keeping the density flat and treating electrons adiabatically. The linear gyrokinetic simulations are conducted with the \texttt{stella} code \citep{barnes2019stella}, of which the details are provided in the supplementary material. The resulting {linear} spectrum is presented in Figure~\ref{fig:stella_simulations}, where the poloidal wavenumber and frequencies have been normalised following the notation in this paper. For comparison to this HSX simulation, an equivalent simulation has also been carried out with a modified geometry in which all field quantities are constant along the field line except the curvature, which is modelled like a single quadratic well {(modelled with parameter $\bar{\omega}_d$ corresponding to the value of bad curvature at the bottom of the central well, and $\Lambda$ as the distance from the centre of the well to the first zero curvature crossing point)}, truncated at a finite good curvature { to prevent spurious modes as discussed later}.
 
\begin{figure}
    \centering
    \includegraphics[width=0.8\textwidth]{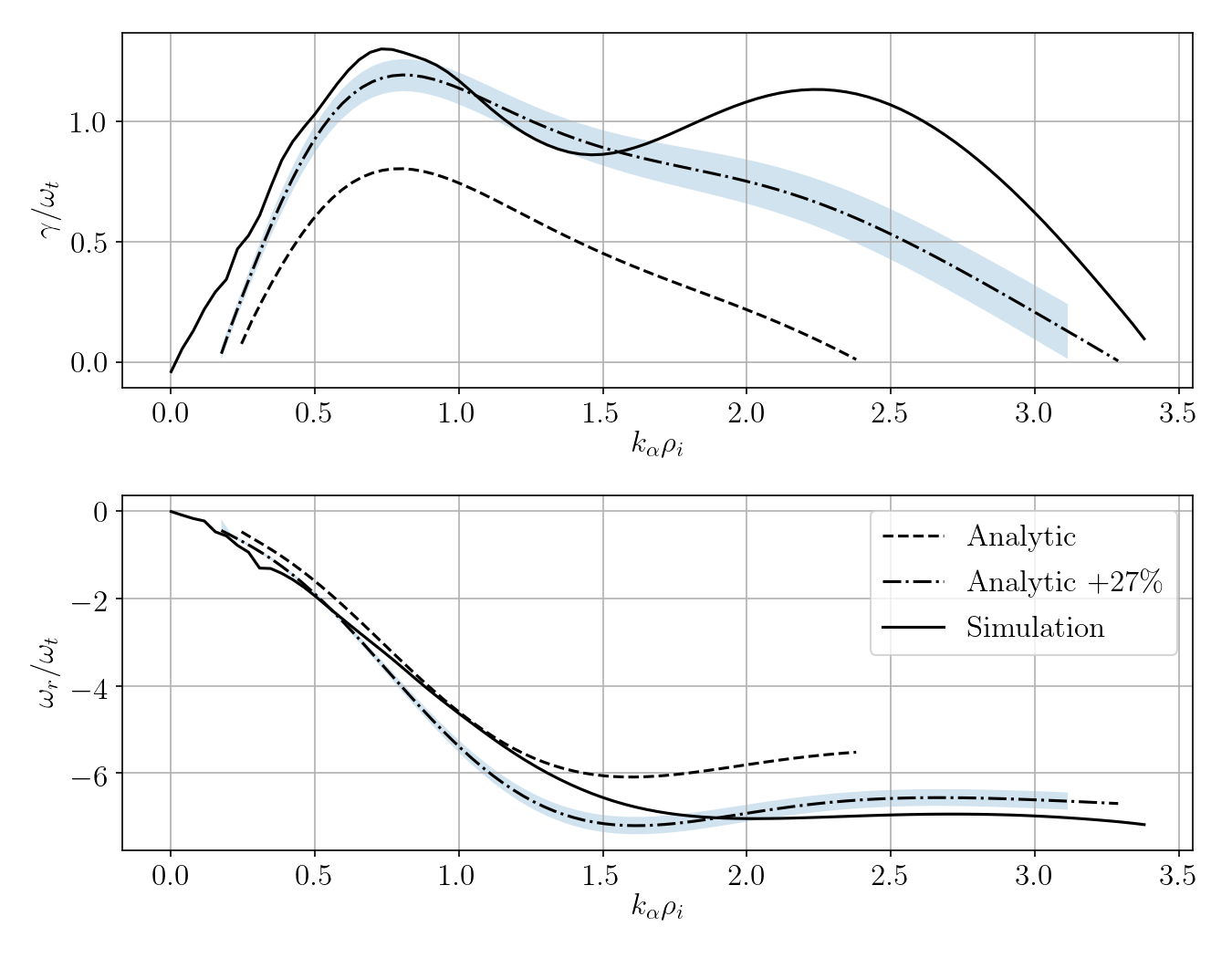}
    \caption{\textbf{Comparison of simulation to analytic model showing quantitative discrepancies.} The plot shows a comparison between the growth rate (top) and frequency (bottom) of the dominant linear ITG mode in the simulation of Figure~\ref{fig:stella_simulations} (solid line) and the analytic model developed in the paper. The dashed line corresponds to the model prediction for $\bar{\omega}_d/\omega_t=1.95$ and $\omega_\star^T/\omega_t=-18$, parameters obtained from the main well of the HSX geometry. The dot-dash line corresponds to the analytical prediction with a $27\%$ larger temperature gradient, with the shade representing $\pm5\%$ variation. This shows that the model suffers as a quantitative predictor. This suggests consideration of the model mainly as a physical qualitative framework to interpret linear spectra behaviour.}
    \label{fig:anal_sim_comparison}
\end{figure}

The first noticeable conclusion from the comparison between these two simulations (solid and dashed lines in Fig.~\ref{fig:stella_simulations}) is that the simplified geometry appears to capture the behaviour of the ITG mode exceptionally well. This strongly backs the approach and geometric approximations considered in this paper, and supports our view on the key role played by the curvature in localising the ITG mode. Thus, we are allowed to describe realistic {linear} stellarator spectra from our theoretical perspective, even though we notice that  the `pure' mode assumption (and perhaps others such as a vigorous drive) does not provide a good quantitative description, in particular requiring somewhat larger temperature gradients in order to match { magnitudes of mode frequency and growth rate of} simulation results. { We show an example of this lack of agreement in Figure~\ref{fig:anal_sim_comparison}, where we include the analytical model prediction for the parameters characterising HSX, and one with a $27\%$ larger temperature gradient.}
\par
We can still investigate the key physical principles we have explored through our model to interpret the {linear} spectrum observed. Let us start from the rightmost part of the spectrum, where we expect to find our $\omega_d$ threshold. In fact, and as shown in Fig.~\ref{fig:stella_simulations} and predicted, this stabilisation point does occur approximately at $\Re\{\omega\}\sim-\bar{\omega}_d$ (numerically within $6\%$). Our interpretation of the nature of this threshold enables us to understand how the spectrum should change as the bad curvature of the field or the temperature gradient are varied. The spectrum would narrow as the curvature is increased or the gradients decreased. Such a simple perspective can explain the narrowing of spectra observed in recent efforts to optimise stellarators for improved temperature gradient thresholds \citep{roberg2022reduction}. In the comparison between the full geometry simulation and that of the reduced geometry, we see that the latter seems to exhibit an additional unstable branch at smaller poloidal wavelengths. This branch corresponds to an anti-ionic temperature driven instability (i.e., rotating in the electronic direction, but not due to kinetic electrons, since their response is adiabatic) that exists in the region of good-curvature of the modified geometry. This behaviour is an interesting case study for the future: if anti-ionic modes localise in good curvature regions, attempting to stabilise the ITG mode by increasing the good curvature in the field could be problematic.
\par
The region beyond the $\omega_d$ threshold in the HSX case considered here does not exhibit other significant instabilities. However, less symmetric geometries such as those in quasi-isodynamic stellarators \citep{podavini2023electrostatic}, often exhibit {linear} spectra with additional structure beyond this point. How can this be framed within our description? That is: how can the ITG mode escape stabilisation by the $\omega_d$ resonance? It is not the localisation of the mode that is suppressing the mode, as it may occur at small $k_\alpha\rho_i$, thus delocalising the mode is not a solution.\footnote{Close to marginality, where the kinetic difference between the $\omega_d$ and Landau damping as explored in previous sections becomes less definite, delocalisation may also become a relevant mechanism even on this side of the spectrum.} The alternative left is for the mode to localise in a different well, a \textit{hopping} mode. Moving to a well that is a priori less unstable because it has better curvature{, or milder FLR effects,} may however be beneficial for the mode because it may be sufficient to make $\Re\{\omega\}\neq-\bar{\omega}_d$ and avoid stabilisation through the $\omega_d$ resonance. Such a jump of the localised mode could occur more than once and we hypothesise, could lead to additional growth rate peaks in the spectrum. These changes on localisation are reminiscent of changes that occur to modes with finite $k_\psi$ \citep{parisi2022three}. { The global behaviour of the localised mode including the possibility of living at different wells could be understood by constructing one spectrum like that in Fig.~\ref{fig:stella_simulations} for each well, with the corresponding change in the model parameters; more precisely, the magnitude of the drift, the well width and the FLR effect (with the latter strongly affecting the scale of $k_\alpha\rho_i$).} The study of this is left for future work, but is an attractive way forward to understanding the behaviour of kinetic ion instabilities. 
\par
Our physics interpretation of the occurrence of the dip in the {linear} spectrum of Fig.~\ref{fig:stella_simulations} from FLR stabilisation and then weakening appears to be also correct. The dotted line in the figure corresponds in fact to the simple expression derived above in Eq.~(\ref{eqn:crit_FLR}). This relative stabilisation of Larmor radius effects can also be observed in the response of the mode { eigenfunction}. 
\begin{figure}
    \centering
    \includegraphics[width = \textwidth]{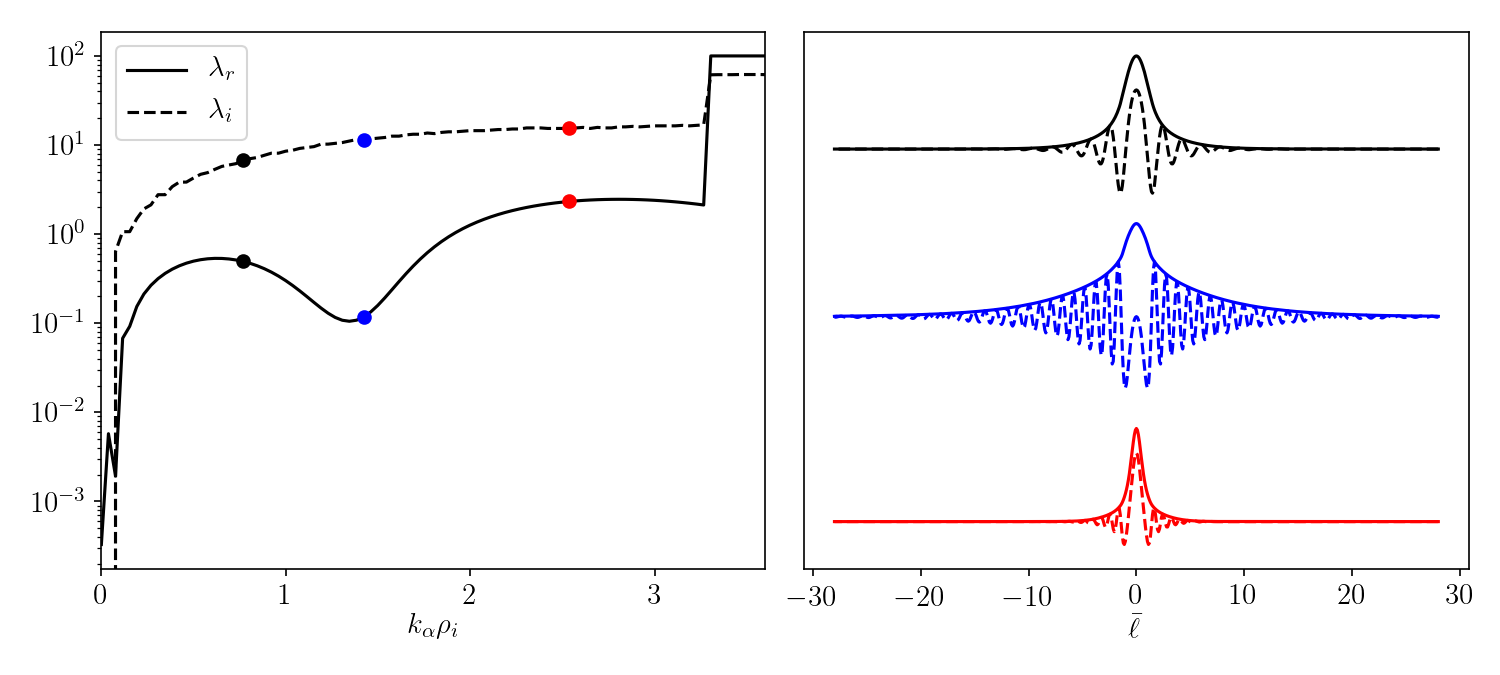}
    \caption{\textbf{Mode structure for HSX gyrokinetic simulations.} The plot shows the parameters $\lambda_r$ and $\lambda_i$ describing the localisation and oscillation of the modes in Fig.~\ref{fig:stella_simulations}. $\lambda_r$ is obtained by fitting a Gaussian $\exp[-\lambda_r\bar{\ell}^2/2]$ to the absolute value of the electrostatic potential $\phi$. The mode is not a pure exponential with generally wider tails. Three examples of the modes in the simplified geometry are provided in the right panel. The $\lambda_i$ parameter is obtained by reading the main oscillation frequency of $\Re\{\phi\}/|\phi|$. Both these measures are inspired by the Gaussian basis used in this paper. The qualitative behaviour observed is fully consistent with the behaviour of our model. That is, $\lambda_i$ monotonically increases with the poloidal wavenumber while the localisation decreases near the Landau threshold ($k_\alpha\rho_i\sim0$), the FLR stabilisation threshold and the drift resonance.}
    \label{fig:stella_simulations_wavefunction}
\end{figure}
We present some example structures from the simulations in Figure~\ref{fig:stella_simulations_wavefunction}. We use the structure from the \texttt{stella} simulations of the simplified geometry for better clarity. The figure also presents a measure of the localisation of the mode, $\lambda_r$, and its structure, $\lambda_i$. Numerically, we compute $\lambda_r$ by fitting a Gaussian $\exp[-\lambda_r\bar{\ell}^2/2]$ to $|\phi|$. Note that the wavefunctions are not really pure Gaussians; in fact, they appear to fit closer to a Gaussian in the centre and a weaker exponential decay further away. Of course the precise behaviour depends on the details of the geometry. Larger $\lambda_r$ denotes a more localised mode. If one takes $\Re\{\phi\}/|\phi|$, the mode then exhibits  a clear periodicity (other than in the middle region). We define $\lambda_i$ to be the main frequency of the that oscillatory behaviour in $\bar{\ell}$. As it is clear in Fig.~\ref{fig:stella_simulations_wavefunction}, as predicted by our treatment, as FLR stabilises the ITG, the mode becomes less localised, while keeping the scale of its mode structure (i.e., $\lambda_i$ unchanged). Note also that the behaviour near the $\omega_d$ resonance gives us some intuition that the `pure' mode form of $\lambda$ does not fully hold there, as the mode only weakly de-localises.
\par
We are then left with the behaviour at the longest poloidal wavelength. As predicted by our theory, the mode becomes increasingly delocalised at lower $k_\alpha\rho_i$. In terms of our Landau threshold picture, this occurs while Landau stabilisation becomes less and less effective. A causality relation between the lack of localisation and growth rate maximisation could perhaps be established  via an energetic argument.  Of course, at some point the model ceases to be valid, and the {linear} spectrum becomes dominated by completely delocalised modes. These very extended modes we may refer to as Floquet or slab-like modes \citep{zocco_xanthopoulos_doerk_connor_helander_2018,zocco_podavini,podavini2023electrostatic}. The difference in growth rate between the two gyrokinetic simulations in Fig.~\ref{fig:stella_simulations} at long wavelengths can be explained to be due to the preeminence of these modes. The presence of such elongated structures make the simulations rather challenging, as long flux tubes are necessary. Their behaviour lies outside the realm of the present model.
\par
We thus understand the very distinct nature of the ITG mode for large and small wavelengths, where the kinetic stabilising mechanisms are different, and thus so is the mode response. This enables one to elucidate the meaning of the {linear} spectrum presented, and offers, as given, a way forward to interpreting {linear} spectra and their physical meaning in more complex geometries.

\section{Conclusions}
In this work, we have proposed a theory of the kinetic Ion-Temperature-Gradient driven mode which features resonant kinetic effects,  with a localised mode structure induced by the field-line-dependent geometry.  We focused on the localising action of the magnetic drift, allowing for general conclusions that apply  both in a tokamak and stellarator  context. 
\par
The magnetic drift spatial dependence models  good and bad curvature regions, introduced with a local well quadratic model. The mathematical description of the problem is based on a power series expansion of the eigenfunctions in the field-following co-ordinate, mitigated by a Gaussian envelope. This generates a hierarchy of coupled eigenvalue problems which for small magnetic drift frequency, strong drive and localised modes, can be truncated. A relatively simple dispersion relation is constructed by considering the simplifying assumption of a singly dominant mode, which we refer to as `pure'. The resulting description features long-wavelength Landau damping, arbitrary Larmor radius effects, and a regularising resonant action of the magnetic drift for short wavelength modes. All these salient physical features are demonstrated to be numerically observed in realistic stellarator geometry, although the simplicity of the model limits  quantitative comparisons with numerical results. Venues for improving the model are also proposed for future work.
\par
The model is also used to provide a prediction for the resonant stabilisation of the toroidal branch of the ITG mode without tacitly assuming constant (along the field line, and thus unrealistic) eigenfunctions. The result is insight into how to tailor the field-line dependence of the magnetic drift in order to suppress the ITG instability. By enforcing that all scales are either Landau damped (at long wavelengths) or kinetically suppressed by the magnetic drift resonance (at short wavelengths) one obtains a critical gradient for ITG destabilisation that scales with $a/L_T\propto 1/(q^{0.6}R).$  This explains why large inverse aspect-ratio devices feature small critical thresholds (as is well known), but also indicates a beneficial effects in having a small safety factor $q$, or more generally short connection length. This final aspect is of particular interest, since it is synergistic with the field-line-bending stabilisation of magnetohydrodynamic (MHD) instabilities \citep{bernstein1958energy,mercier1962,greene1962,correa1978,CHT}. These expectations, scalings, synergies and behaviour alongside MHD stability will be the subject of careful analysis and simulation in a future paper \citep{Edu_AZ}.

\section*{Data availability}
The data that support the findings of this study are openly available at the Zenodo repository with DOI/URL 10.5281/zenodo.11388974.

\section*{Acknowledgements}
We gratefully acknowledge fruitful discussion with R. Nies, P. Costello, G. Plunk, G. Roberg-Clark, L. Podavini and F. Parra. 

\section*{Funding}
E. R. was supported by a grant by Alexander-von-Humboldt-Stiftung, Bonn, Germany, through a postdoctoral research fellowship. Part of this work was conceived during the Simons Collaboration on Hidden Symmetries meetings, to which we are grateful for its support.  

\section*{Declaration of interest}
The authors report no conflict of interest.

\appendix

\section{Taylor-Gauss expansion of GK equation} \label{app:exp_HG_GK}
In this Appendix we detail the Taylor-Gauss expansion of the GK equation. Our starting point is the GK equation, Eq.~(\ref{eqn:GK_schrod_herm}), which we write as, 
\begin{equation}
    2\left(\frac{\omega_tx_\parallel}{\omega}\right)^2\partial_{\bar{\ell}}^2 g+\left(1-\frac{\tilde{\omega}_d}{\omega}(\bar{\ell}^2-1)\right)g+4\tilde{\omega}_d\left(\frac{\omega_tx_\parallel}{\omega}\right)^2\bar{\ell}\partial_{\bar{\ell}}g=\frac{q_i}{T_i}F_{0i}J_0\left(1-\frac{\tilde{\omega}_\star}{\omega}\right)\phi, \tag{\ref{eqn:GK_schrod_herm}}
\end{equation}
To obtain our Taylor-Gauss resolution of the equation we then substitute,
\begin{equation}
    g(\bar{\ell},\mathbf{v})=\sum_{n=0}^\infty g_n(\mathbf{v})\sqrt{\frac{\Re\{\lambda\}^n}{n!}}\bar{\ell}^n e^{-\lambda\bar{\ell}^2/2}, \tag{\ref{eqn:g_exp_herm}}
\end{equation}
and likewise for $\phi$, into the equation. Note that the normalisation is chosen here in such a way that the magnitude of the basis (i.e., the terms multiplying $g_n$) are considered roughly of order one.
\par
The action of the second derivative on this basis yields,
\begin{align}
    \partial_{\bar{\ell}}^2 g(\bar{\ell},\mathbf{v})=&\sum_{n=0}^\infty g_ne^{-\lambda\bar{\ell}^2/2}\sqrt{\frac{\Re\{\lambda\}^n}{n!}}\left[n(n-1)\bar{\ell}^{n-2}-(2n+1)\lambda\bar{\ell}^n+\lambda^2\bar{\ell}^{n+2}\right] \nonumber \\
    =& \sum_{n=0}^\infty\frac{\lambda\bar{\ell}^n}{N(n)}e^{-\lambda\bar{\ell}^2/2}\left[\sqrt{(n+2)(n+1)}\frac{\Re\{\lambda\}}{\lambda}g_{n+2}-(2n+1)g_n+\sqrt{n(n-1)}\frac{\lambda}{\Re\{\lambda\}}g_{n-2}\right] . \label{eqn:d2_TayGau}
\end{align}
In this notation it should be interpreted that $g_n=0$ for $n<0$, and likewise any negative power of $\bar{\ell}$ in the summation should be taken to be zero. 
\par
For the terms that present a product with the drift frequency, and thus with a second power of $\bar{\ell}^2$, we simply have,
\begin{equation}
    \bar{\ell}^2 g(\bar{\ell},\mathbf{v})=\sum_{n=0}^\infty\bar{\ell}^ne^{-\lambda\bar{\ell}^2/2}\sqrt{\frac{\Re\{\lambda\}^n}{n!}}\frac{1}{\Re\{\lambda\}}\sqrt{n(n+1)}g_{n-2}, \label{eqn:l2_TG}
\end{equation}
and,
\begin{equation}
    \bar{\ell}\partial_{\bar{\ell}}g=\sum_{n=0}^\infty\bar{\ell}^ne^{-\lambda\bar{\ell}^2/2}\sqrt{\frac{\Re\{\lambda\}^n}{n!}}\left(n g_n-\frac{\lambda}{\Re\{\lambda\}}\sqrt{n(n-1)}g_{n-2}\right).
\end{equation}
{ Similar expressions would be obtained if a Hermite basis was used instead.}
With these expressions and collecting terms, we get the general equation 
\begin{multline}
    E_n=2\Re\{\lambda\}\left(\frac{\omega_t x_\parallel}{\omega}\right)^2\sqrt{(n+2)(n+1)}g_{n+2} + \\
    +\left[1+\frac{\tilde{\omega}_d}{\omega} -2\lambda(2n+1)\left(\frac{\omega_tx_\parallel}{\omega}\right)^2+4n\frac{\tilde{\omega}_d}{\omega}\left(\frac{\omega_t x_\parallel}{\omega}\right)^2\right]g_n+\\
    +\frac{2}{\Re\{\lambda\}}\left[\lambda^2\left(\frac{\omega_t x_\parallel}{\omega}\right)^2-\frac{\tilde{\omega}_d}{2\omega}-2\lambda\frac{\tilde{\omega}_d}{\omega}\left(\frac{\omega_t x_\parallel}{\omega}\right)^2\right]\sqrt{n(n-1)}g_{n-2}-\\
     -\frac{q_i}{T_i}F_{0i}J_0\left(1-\frac{\tilde{\omega}_\star}{\omega}\right)\phi_n.
    \tag{\ref{eqn:GK_schrod_herm_n}}
\end{multline}
presented in the main text, and to be interpreted as,
\begin{equation}
    \sum_{n=0}^\infty E_{n}\bar{\ell}^ne^{-\lambda\bar{\ell}^2/2}\sqrt{\frac{\Re\{\lambda\}^n}{n!}}=0.
\end{equation}
It is clear that for an exact solution to the system we need to satisfy $\{E_n=0\}$ for all $n$, if the equation is to be satisfied for all $\bar{\ell}$. 

\section{Details on dispersion construction and higher order modes} \label{app:high_n_considerations}
In this Appendix we show the details on how the construction of the dispersion function in the text is done, including the generalisation to other modes other than the simplest Gaussian $n=0$. 
\subsection{General structure of the problem}
To obtain the dispersion relation of whichever mode we are interested in investigating, it is necessary to construct the matrix $\mathbb{D}$ in Eq.~(\ref{eqn:gn_Dnm_phim}). That is, we must make the appropriate combinations of the coefficients in Eq.~(\ref{eqn:GK_schrod_herm_n}) to bring the set of equations to the form considered in Eq.~(\ref{eqn:gn_Dnm_phim}). Because we are dealing with a system of equations with dimension $N$, our truncation number, we would like some systematic way in which to construct the relevant entries in the matrix. In particular, we would like to exploit our ordering in $\delta$ and $\epsilon$ to simplify the procedure before explicitly constructing the equations. 
\par
To that end, let us write the system of equations defined by the first $N/2+1$ equations in matrix form,
\begin{equation}
    \mathsfbi{G}_{ij}\mathbf{g}_j=\pmb{\Phi}_{ij}\pmb{\phi}_j,
\end{equation}
where the vectors $\mathbf{g}$ and $\pmb{\phi}$ contain the modes from $n=0$ to $N$, and the matrices can be constructed following the general expression for $E_n$ in Eq.~(\ref{eqn:GK_schrod_herm_n}). Because the system considers separately the even and odd orders, we shall restrict these matrices to the even or odd parts of the problem separately. The arguments to follow are independent of which class we consider (with minor changes), but we choose the even part of the solution (just because one must be chosen).
\par
Because we are about to invoke some ordering considerations to simplify the system of equations, we shall fairly represent the order of each mode. We shall treat first each $g_n$ and $\phi_n$ in an unordered fashion, and thus only consider the ordering of the factors we have explicitly in Eq.~(\ref{eqn:GK_schrod_herm_n}). For simplicity, and in hindsight of what will end up later being considered, we shall take the following consistent scaling for $\lambda\sim\sqrt{\delta}(\omega/\omega_t)$. By itself it is not ordered in any particular way (it simply cannot be too small), but $\lambda(\omega_t/\omega)^2\sim O(\epsilon)$ and $\lambda^2(\omega_t/\omega)^2\sim O(\delta)$. With this in mind it is straightforward to picture the ordering of each element in the matrices,
\begin{equation}
    \mathsfbi{G}=
    \begin{pmatrix}
        1 & \epsilon &  &  &\\
        \epsilon  & 1 & \epsilon &  & \\
         & \epsilon  & 1 & \epsilon &  \\
         &  &  \ddots & \ddots & \ddots \\
         &  & & \epsilon & 1 & \epsilon \\
         &  & & & \epsilon & 1 \\
    \end{pmatrix}, ~ ~~~  \pmb{\Phi}=
    \begin{pmatrix}
        1 &  &  &  &\\
          & 1 &  &  & \\
         &   & 1 &  &  \\
         &  &  & \ddots &  \\
         &  & &  & 1 &  \\
         &  & & &  & 1 \\
    \end{pmatrix}.
\end{equation}
We note that $\mathsfbi{G}$ is a tridiagonal matrix while $\pmb{\Phi}$ only has non-zero entries in the main and lower diagonals. And importantly, hopping off-diagonal terms are ordered like $\epsilon$, which shall allow us to simplify the problem significantly. To order these matrix elements the way we have done, we must note that the hopping terms as considered do not only bring $\epsilon$, but are also proportional to the mode number $n$. Thus, to preserve the ordering $\epsilon$ of the off-diagonals we shall limit the truncation $N\ll N_\epsilon=1/\epsilon$.
\par
With these matrices set up, we may now attempt the approximate construction of $\mathbb{D}=\mathsfbi{G}^{-1}\pmb{\Phi}$ to the right order $O(\epsilon^2)$. Fortunately, the inversion of a tridiagonal matrix can be expressed succinctly in the following form by \cite{usmani1994inversion}. Defining the elements along the main diagonal as $a_i$ from $i=1,\dots,N/2$,\footnote{If we were considering the truncated system for the odd parts, then we would end at $(N+1)/2$.} and the upper and lower diagonals as $b_i$ and $c_i$ respectively (also starting at $n=1$) for matrix $\mathsfbi{G}$, the inverse may be written as,
\begin{equation}
    (\mathsfbi{G}^{-1})_{ij}=\begin{cases}
        (-1)^{i+j}b_i\dots b_{j-1}\theta_{i-1}\phi_{j+1}/\theta_n & (i\leq j) \\
        (-1)^{i+j}c_j\dots c_{i-1}\theta_{j-1}\phi_{i+1}/\theta_n & (i> j),
    \end{cases} \label{eqn:Usmani_tridiag}
\end{equation}
where $\theta_i=a_i\theta_{i-1}-b_{i-1}c_{i-1}\theta_{i-1}$ with $\theta_0=1$ and $\theta_1=a_1$, and $\phi_i=a_i\phi_{i+1}-b_ic_i\phi_{i+2}$ with $\phi_{N/2+1}=1$ and $\phi_{N/2}=a_{N/2}$. Now, recall we are interested in the construction to order $O(\epsilon)$, and thus, we may drop any term that is higher order. In particular, this implies dropping in the iteration expressions the terms involving products of $b$ and $c$, which as we indicated above, are each order $\epsilon$. As a result, $\theta_i=a_i\dots a_1$ and $\phi_i=a_{N/2}\dots a_i$. Finally, restricting the products of $b$ and $c$ involved in Eq.~(\ref{eqn:Usmani_tridiag}) not to surpass the right order, we may write the inverse succinctly as follows,
\begin{equation}
    (\mathsfbi{G}^{-1})_{ij}=\frac{1}{a_i}\delta_{ij}-\frac{b_i}{a_ia_{i+1}}\delta_{i,j-1}-\frac{c_{i-1}}{a_ia_{i-1}}\delta_{i,j+1}.
\end{equation}
Define now the diagonal elements of $\pmb{\Phi}_{ij}=p_j\delta_{ij}$; the matrix product is then, to leading order,
\begin{equation}
    \mathbb{D}_{2i,2j}=\left(\mathsfbi{G}^{-1}\pmb{\Phi}\right)_{ij}=\frac{p_i}{a_i}\delta_{ij}-\frac{c_jp_j}{a_ia_j}\delta_{i-1,j}-\frac{b_ip_j}{a_ia_j}\delta_{i+1,j}.
\end{equation}
To further simplify, let us be more explicit on the various coefficients of $\mathsfbi{G}$ and $\pmb{\Phi}$. Reading the expressions off Eq.~(\ref{eqn:GK_schrod_herm_n}) to the correct order,
\begin{subequations}
    \begin{align}
        p_{n/2}&=\frac{q_i}{T_i}F_{0i}J_0\left(1-\frac{\tilde{\omega}_\star}{\omega}\right), \\
        a_{n/2}&\approx 1+\frac{\tilde{\omega}_d}{\omega}-2\lambda(2n+1)\left(\frac{\omega_t x_\parallel}{\omega}\right)^2, \\
        b_{n/2}&=2\Re\{\lambda\}\left(\frac{\omega_t x_\parallel}{\omega}\right)^2\sqrt{(n+2)(n+1)}, \\
        c_{n/2-1}&=\frac{2}{\Re\{\lambda\}}\sqrt{n(n-1)}\left[\lambda^2\left(\frac{\omega_t x_\parallel}{\omega}\right)^2-\frac{\tilde{\omega}_d}{2\omega}\right].
    \end{align}
\end{subequations}
With these, we may write, at once,
\begin{subequations}
    \begin{align}
        \mathbb{D}_{nn}&=\frac{q_i}{T_i}F_{0i}J_0\left(1-\frac{\tilde{\omega}_\star}{\omega}\right)\frac{1}{1+\frac{\tilde{\omega}_d}{\omega}-2\lambda(2n+1)\left(\frac{\omega_t x_\parallel}{\omega}\right)^2}, \label{eqn:Dnn}\\
        \mathbb{D}_{n,n-2}&=-\frac{q_i}{T_i}F_{0i}J_0\left(1-\frac{\tilde{\omega}_\star}{\omega}\right)\frac{2}{\Re\{\lambda\}}\sqrt{n(n-1)}\left[\lambda^2\left(\frac{\omega_t x_\parallel}{\omega}\right)^2-\frac{\tilde{\omega}_d}{2\omega}\right], \label{eqn:Dnn-2}\\
        \mathbb{D}_{n,n+2}&=-\frac{q_i}{T_i}F_{0i}J_0\left(1-\frac{\tilde{\omega}_\star}{\omega}\right)2\Re\{\lambda\}\sqrt{(n+1)(n+2)}\left(\frac{\omega_t x_\parallel}{\omega}\right)^2,\label{eqn:Dnn+2}
    \end{align}
\end{subequations}
as the only relevant matrix elements to order $\epsilon$. 
\par
With this matrix in place, we are in a position to apply quasineutrality to construct the system of equations on $\phi_n$ in Eq.~(\ref{eqn:sys_eq_GK_n}). Let us be explicit in the construction of this system by writing,
\begin{equation}
    \sum_{j=0}^N \mathbb{M}_{ij}\phi_j=0, \tag{\ref{eqn:sys_eq_GK_n}}
\end{equation}
and write with $\mathcal{D}_{nm}=(1/\bar{n})\int\mathrm{d}^3\mathbf{v}J_0\mathbb{D}_{nm}$,
\begin{equation}
    \mathbb{M}_{nm}=(1+\tau-\mathcal{D}_{nn})\delta_{nm}-\mathcal{D}_{n,n-2}\delta_{n-2,m}-\mathcal{D}_{n,n+2}\delta_{n+2,m}.
\end{equation}

\subsection{Solving for a `pure' $M$-th mode}
Let us now focus on the problem when there is a dominant $M$-th mode, with $M\ll N$. That is, let us take $\phi_M\sim O(1)$. In addition we make the choice $\phi_n=0$ for $n>M$, with which we deem the description of a `pure' mode (more comments on this to follow later). We may then write the equations one by one starting from the $M$-th mode down,
\begin{align*}
    -\mathcal{D}_{M+2,M}\phi_M&=0, \\
    (1+\tau-\mathcal{D}_{MM})\phi_M-\mathcal{D}_{M,M-2}\phi_{M-2}&=0, \\
    (1+\tau-\mathcal{D}_{M-2,M-2})\phi_{M-2}-\mathcal{D}_{M-2,M-4}\phi_{M-4}-\mathcal{D}_{M-2,M}\phi_{M}&=0, \\
    \vdots    
\end{align*}
This system of equations can be straightforwardly solved to $O(\epsilon)$ by taking the following ordering for the various modes of $\phi$: $\phi_M\sim O(1)$, $\phi_{M-2}\sim O(\epsilon)$ and $\phi_{M-2k}\sim O(\epsilon^k)$. In that case, the consistent solution to the problem is the following dispersion relation condition on $\omega$ and $\lambda$,
\begin{subequations}
    \begin{align}
        1+\tau-\mathcal{D}_{MM}=0,   \label{eqn:dispRelMM}\\
        \mathcal{D}_{M+2,M}=0,
    \end{align} \label{eqn:dispRelM}
\end{subequations}
together with
\begin{equation}
    \phi_{M-2}=\frac{\mathcal{D}_{M-2,M}}{1+\tau-\mathcal{D}_{M-2,M-2}},
\end{equation}
which is order $\epsilon$. Thus for studying the $M$-th mode, we must solve Eqs.~(\ref{eqn:dispRelM}). 
\par
Let us start first by investigating the second condition, namely $\mathcal{D}_{M+2,M}=0$. Dropping unimportant factors from Eq.~(\ref{eqn:Dnn-2}), we are left with the integral condition,
\begin{equation}
    \int\mathrm{d}^3\mathbf{v}J_0^2 e^{-v^2/v_{Ti}^2}\left(1-\frac{\tilde{\omega}_\star}{\omega}\right)\left[\lambda^2\left(\frac{\omega_t x_\parallel}{\omega}\right)^2-\frac{\tilde{\omega}_d}{2\omega}\right]=0. \label{eqn:lam_M_integral_eq}
\end{equation}
To perform the integrals over velocity space, we use $\tilde{\omega}_\star=\omega_\star\left[1+\eta(x_\parallel^2+x_\perp^2-3/2)\right]$ and $\tilde{\omega}_d=\bar{\omega}_d(x_\parallel^2+x_\perp^2/2)$, and for simplicity, drop the FLR contributions to the integral, so that,
\begin{equation}
    \lambda=\sqrt{\frac{\omega\bar{\omega}_d}{\omega_t^2}}.
\end{equation}
This is the characteristic localisation of the $M$-th mode. Note that it is actually $M$-independent, meaning that the differences between modes arise from the other expression in Eq.~(\ref{eqn:dispRelM}). We shall refer to this equation as the $M$-th mode dispersion relation, and we may explicitly write it as,
\begin{equation}
    \mathcal{D}=1+\tau+\frac{\zeta}{\bar{n}}\int J_0^2\left(1-\frac{\tilde{\omega}_\star}{\omega}\right)\frac{f_0}{x_\parallel^2-\zeta\left(1+\frac{\bar{\omega}_d}{2\omega}x_\perp^2\right)}\mathrm{d}^3\mathbf{v}, \tag{\ref{eqn:disp_fun_kin_itg}}
\end{equation}
where,
\begin{equation}
    \zeta=\frac{1}{2}\left[\lambda(2M+1)\left(\frac{\omega_t}{\omega}\right)^2-\frac{\bar{\omega}_d}{2\omega}\right]^{-1}. \label{eqn:zeta_def_M_or}
\end{equation}
for any $0\leq M\ll N$, and both for even and odd $M$ without distinction. The mode number solely enters the problem through the equations in the kinetic parameter $\zeta$, and in particular, in its streaming contribution. The larger the mode number, the larger the contribution from the streaming term. 
\par
Note that one could argue that this particular `pure' solution to the problem is not the only one. In particular, and following the ordering argument of $\phi_M\sim O(1)$ and $\phi_{M-2}\sim O(\epsilon)$, one could consider constructing a solution in a more symmetric way around the $M$-th mode, where the $\phi_n$ for $n>M$ are not exactly zero, but ordered like $O(\epsilon^{(n-M)/2})$. In that instance, the description to order $\epsilon$ would leave us with a single dispersion equation, namely Eq.~(\ref{eqn:dispRelMM}). Thus, even if $\lambda$ should at least have an ordering like that in the `pure' mode, its precise form would not be constrained. In what sense is then the `pure' mode an illustrative choice? There are two important arguments to defend the preeminence of the `pure' mode, albeit not fully conclusive. The first, is that by making the choice of $\lambda$ above, and as explicitly shown, we make, to order $\epsilon$, the lower off-diagonal terms of $\mathcal{D}$ vanish. As such, the connection of modes to higher harmonics is broken, in a sort of closure scheme, preventing problems with factors of increasing mode numbers and isolating the solution from the truncation point. Second of all, this choice of $\lambda$ leads to an agreement of our system with the fluid limit in the limit of the latter. Other choices of $\lambda$ would not do so. And finally, it is the simplest choice to make. All this invites us to study the `pure' modes in this paper. However, in doing so we expect to find discrepancies with the full problem, which likely involves a more subtle involvement of $\lambda$. Evidence of this is shown in the numerical comparison to simulations. Although many of the qualitative features of the spectra can be well captured and explained, predicting the exact form and dependence a priori is out of reach. Although this treatment of $\lambda$ is the weakest point of the treatment, understanding the `pure' mode behaviour is highly insightful. Future work may be devoted to improving the model by perhaps allowing $\lambda$ as a free parameter to optimise to extremise the growth rate of the ITG (much like a ballooning parameter).

\subsubsection{FLR corrections to $\lambda$}
\begin{figure}
    \centering
    \includegraphics[width = 0.6\textwidth]{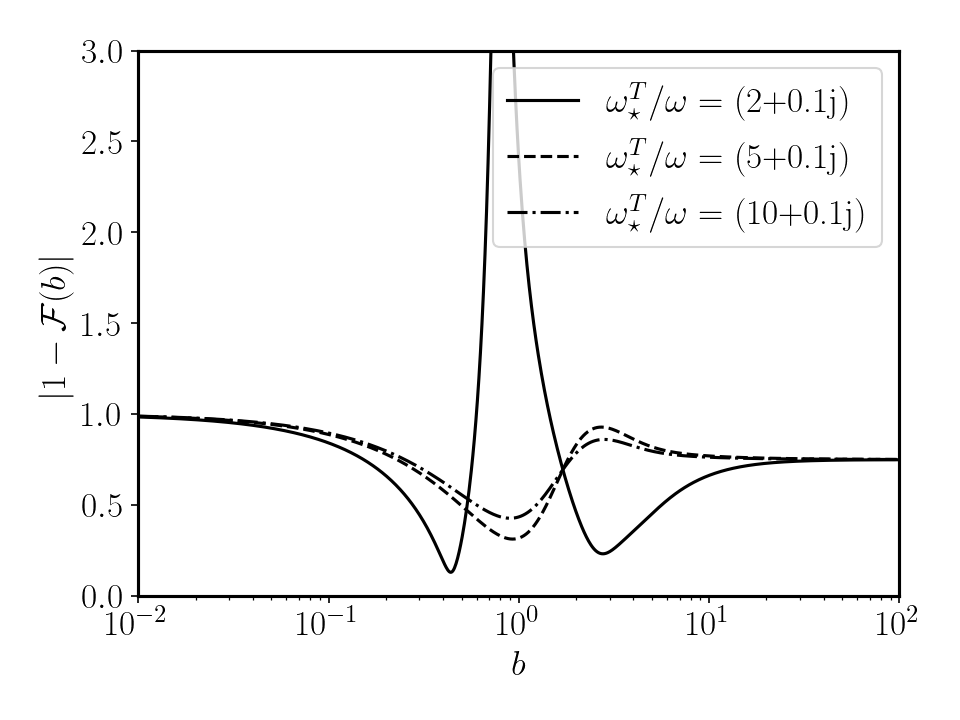}
    \caption{\textbf{Finite Larmor radius corrections to $\lambda$.} Correction factor to $\lambda$ due to finite Larmor radius effects for a number of different temperature gradient drives, $\omega_\star^T/\omega$, for an ionic unstable mode. The effects of FLR on $\lambda$ are moderate at large diamagnetic drive, but become very significant near $b\sim1$ at lower drives.}
    \label{fig:FLR_corr_lam}
\end{figure}

To give a flavour of variations that $\lambda$ may be subject to even within the `pure' mode framework, let us show how to include the FLR corrections in Eq.~(\ref{eqn:lam_M_integral_eq}). The full-FLR form of the localisation parameter $\lambda$ takes the form,
\begin{equation}
    \lambda=\sqrt{\frac{\omega\bar{\omega}_d}{\omega_t^2}\left[1-\mathcal{F}(b)\right]},
\end{equation}
where 
\begin{equation}
    \mathcal{F}(b)=\frac{b}{2}\frac{(\Gamma_0-\Gamma_1)\left(1-\frac{\omega_\star}{\omega}+2\frac{\omega_\star^T}{\omega}(b-1)\right)-\frac{\omega_\star^T}{\omega}\Gamma_1}{\Gamma_0\left(1-\frac{\omega_\star}{\omega}+\frac{\omega_\star^T}{\omega}(b-1)\right)-b\frac{\omega_\star^T}{\omega}\Gamma_1}.
\end{equation}
It follows from this form that, indeed, in the limit of a small FLR, $\mathcal{F}\rightarrow0$, and thus we recover the simple limit $\lambda\sim\sqrt{\omega\bar{\omega}_d/\omega_t^2}$. In the opposite limit, we find that $\lambda\approx (\sqrt{3}/2) \lambda_{b=0}$, which is roughly a $\sim13\%$ reduction; i.e., the longitudinal scale of the mode will increase by roughly a $\sim7\%$ with respect to the no-FLR expectation. The maximum deviation tends to occur when $b\sim1$, corresponding to the point where the FLR effects are most efficient. At moderate values for $\omega_\star^T/\omega$ note that the amplification can be significant following the potentially resonant denominator (see Figure~\ref{fig:FLR_corr_lam}). In the strong drive limit the simple `pure' mode should nevertheless be an appropriate qualitative description even if we do not include the variation of $\lambda$. There would be no point in keeping a significant added complication to a description through the `pure' mode which is already a convenience choice. 

\section{Expressing the dispersion in terms of the dispersion function} \label{app:plasma_disp_function}
In this appendix we detail the construction of the final form of the dispersion function $\mathcal{D}$. We start from,
\begin{equation}
    \mathcal{D}=1+\tau+\frac{\zeta}{\bar{n}}\int J_0^2\left(1-\frac{\tilde{\omega}_\star}{\omega}\right)\frac{f_0}{x_\parallel^2-\zeta\left(1+\frac{\bar{\omega}_d}{2\omega}x_\perp^2\right)}\mathrm{d}^3\mathbf{v},\tag{\ref{eqn:disp_fun_kin_itg}}
\end{equation}
and recall from the main text that we first considered the integral over $x_\parallel$. In doing such an integral, we can write the problem in terms of plasma dispersion functions. However, given the form of $I_{\parallel,\beta}$, Eq.~(\ref{eqn:I_par_beta}), we need to spell out the powers of $x_\parallel$ in the problem explicitly. To do so we must recall the definitions of $\tilde{\omega}_d$ and $\tilde{\omega}_\star$. With that, the numerator of the integrand,
\begin{align}
    \left(1-\frac{\tilde{\omega}_\star}{\omega}\right)=&\left[1-\frac{\omega_\star}{\omega}+\frac{\omega_\star^T}{\omega}\left(\frac{3}{2}-x_\perp^2\right)\right]-\frac{\omega_\star^T}{\omega}x_\parallel^2\nonumber\\
    &= \mathcal{A}+\mathcal{B}x_\parallel^2.
\end{align}
With this $x_\parallel$ dependence made explicit, we may perform the first integral over $x_\parallel$ in Eq.~(\ref{eqn:disp_fun_kin_itg}). Writing the velocity space measure explicitly as,
\begin{equation}
    \mathrm{d}^3\mathbf{v}=2\pi(v_T)^3x_\perp\mathrm{d}x_\perp\mathrm{d}x_\parallel,
\end{equation}
we have,
\begin{align}
    I&=\frac{1}{\bar{n}}\int f_0J_0^2\frac{\left(1-\frac{\tilde{\omega}_\star}{\omega}\right)}{x_\parallel^2-\zeta\left(1+\frac{\bar{\omega}_d}{2\omega}x_\perp^2\right)}\mathrm{d}^3\mathbf{v}= \label{eqn:integral_disp_x_perp}\\
    &=2\int_0^\infty J_0\left(x_\perp\sqrt{2b}\right)^2 x_\perp e^{-x_\perp^2}\left[\mathcal{A}Z_0(\bar{\zeta})+\mathcal{B}Z_1(\bar{\zeta})\right]
\end{align}
where,
\begin{gather}
    \bar{\zeta}=\zeta\left(1+\frac{\bar{\omega}_d}{2\omega}x_\perp^2\right), \\
    Z_0(x)=\frac{Z(\sqrt[*]{x})}{\sqrt[*]{x}},\\
    Z_1(x)=\left(1+\sqrt[*]{x} Z(\sqrt[*]{x})\right), \\
    Z_2(x)=\frac{1}{2}\left[1+2x(1+\sqrt[*]{x}Z(\sqrt[*]{x}))\right].
\end{gather}
It is now the time of performing the integral over $x_\perp$. The main difficulty here is that $x_\perp$ appears in the arguments of the plasma dispersion functions over which we need to be integrating. Integrals of this form for small FLR effects have been recently found in exact form by \cite{ivanov2023analytical}, but full FLR effects are sought here. To that end, we proceed by exploiting the $\bar{\omega}_d/\omega\ll1$ assumption, together with the exponential $\exp[-x_\perp^2]$ that limits the values of $x_\perp$ from becoming much larger than one, and expand the functions in the integrand. Application of Taylor expansion and the chain rule will yield different powers of $x_\perp$ in the integrand. This procedure is straightforward, and may be efficiently calculated with the aid of computer algebra. Note that by performing this expansion, we are losing the $\omega_d$ resonance in $x_\perp$ while we have kept its $x_\parallel$ part. The $\omega_d$ resonance effects are thus only partially captured.
\par
Integrals over powers of $x_\perp$ with the other factors in the integrand of Eq.~(\ref{eqn:integral_disp_x_perp}) are well known \citep{kadomtsev1970turbulence}\citep[Eq.~6.615]{gradshteyn2014table}. Thus, all we need to do is collect powers of $x_\perp$ and collect terms. Terms corresponding to a particular power of $x_\perp$ will be multiplied by the appropriate FLR factor that results from the integral. The relevant $x_\perp$ integrals are,
\begin{subequations}
\begin{gather}
    F_0(b)=2\int_0^\infty J_0\left(x_\perp\sqrt{2b}\right)^2 x_\perp e^{-x_\perp^2}=\Gamma_0(b), \\
    F_2(b)=2\int_0^\infty J_0\left(x_\perp\sqrt{2b}\right)^2 x_\perp^3 e^{-x_\perp^2}=(1-b)\Gamma_0(b)+b\Gamma_1(b), \\
    F_4(b)=2\int_0^\infty J_0\left(x_\perp\sqrt{2b}\right)^2 x_\perp^5 e^{-x_\perp^2}=2\left[(1-b)^2\Gamma_0(b)+\left(\frac{3}{2}-b\right)b\Gamma_1(b)\right], \\
\end{gather}
% \begin{multline}
%         F_6(b)=2\int_0^\infty J_0\left(x_\perp\sqrt{2b}\right)^2 x_\perp^7 e^{-x_\perp^2}=\\
%         =[6-(b-2)b(4b-9)]\Gamma_0(b)+b(b-1)(4b-11)\Gamma_1(b).
% \end{multline}\label{eqn:weber_integrals}
\end{subequations}
The resulting $\mathcal{D}$ can be written as,
\begin{equation}
    \mathcal{D}=1+\tau+\sum \mathcal{T}_{(\alpha,\beta,\gamma)}\frac{\bar{\omega}_d^\alpha\omega_\star^\beta}{\omega^{\alpha+\beta}}F_\gamma(b).
\end{equation}
Some of the leading order $\mathcal{T}_{(\alpha,\beta,\gamma)}$ terms are shown in Table~\ref{tab:terms_disp_relation} for reference. Many of the terms included are not necessary. They are not with regards to explaining the physical behaviour of the mode, and are in addition higher order in $\delta$ and $\epsilon$ than originally devised for. Nevertheless, it may be helpful in analysing the behaviour of the dispersion equation. 

\begin{table}
    \centering
    \begin{tabular}{c|c|c||c}
        $\alpha$ & $\beta$ & $\gamma$ & $\mathcal{T}_{(\alpha,\beta,\gamma)}$   \\ \hline
        0 & 0 & 0 & $\sqrt{\zeta} Z(\sqrt{\zeta})$   \\
        0 & 1 & 0 & $~~-\left(1-\frac{3\eta}{2}\right)\sqrt{\zeta}Z(\sqrt{\zeta})-\eta\zeta Z_+(\sqrt{\zeta})~~$ \\
        0 & 1 & 2 & $~~-\eta \sqrt{\zeta} Z(\sqrt{\zeta})~~$
        \\\hline
        1 & 0 & 2 & $~~\frac{1}{4}\left[1-(1+2\zeta)Z_+(\sqrt{\zeta})\right]~~$
        \\
        1 & 1 & 2 & $~~\frac{1}{4}\left\{-1+\eta\left(\frac{3}{2}+\zeta\right)+\left[1+2\zeta+\eta\left(-\frac{3}{2}+2\zeta(\zeta-2)\right)\right]Z_+(\sqrt{\zeta})\right\}~~$
        \\
        1 & 1 & 4 & $~~\frac{\eta}{4}\left[-1+(1-2\zeta)Z_+(\sqrt{\zeta})\right]~~$ 
        \\\hline
        2 & 0 & 4 & $~~\frac{1}{16}\left[-\frac{3}{2}+\zeta+\left(\frac{3}{2}+2\zeta(1+\zeta)\right)Z_+(\sqrt{\zeta})\right]~~$ 
        \\
        2 & 1 & 4 & $~~\frac{1}{32}\left\{3-2\zeta-\eta\left(\frac{9}{2}+2\zeta(\zeta-1)\right)+\left[-3-4\zeta(1+\zeta)+\eta\left(\frac{9}{2}+\zeta\left(7+2\zeta(5-2\zeta)\right)\right)\right]Z_+(\sqrt{\zeta})\right\}~~$ 
        \\
        2 & 1 & 6 & $~~\frac{\eta}{16}\left[\frac{3}{2}-\zeta-\left(\frac{3}{2}+2\zeta(1+\zeta)\right)Z_+(\sqrt{\zeta})\right]~~$ 
        \\\hline        
        
    \end{tabular}
    \caption{\textbf{Contributions to the dispersion relation.} Different term contributions to the dispersion relation $\mathcal{D}$ which may or may not be included depending on the required approximation. The columns $\alpha$ and $\beta$ denote the powers of $\bar{\omega}_d/\omega$ and $\omega_\star/\omega$ respectively that these terms are multiplied by, Column $\gamma$ labels the FLR function that is multiplied by these terms, related directly to the number of powers of $x_\perp$ prior to integrating over $x_\perp$. The notation $Z_+(\sqrt{\zeta})=1+\sqrt{\zeta}Z(\sqrt{\zeta})$ for brevity.  }
    \label{tab:terms_disp_relation}
\end{table}

\section{Full Larmor radius form of the ITG fluid equation} \label{app:full_flr_fluid}
In this Appendix we sketch the derivation of the full-Larmor-radius form of the ITG fluid equation. We follow closely the work of \cite{connor1980stability}, and where possible we shall simply quote this work. Let us remind ourselves about the set-up of the fluid ITG problem. We start by assuming that the transit time is long compared to the characteristic time of the instability ($\omega_t/\omega\sim\epsilon \ll1$), so that we may consider expanding our solution to the GK equation ignoring any kinetic resonance there.
\par
The general solution to the GK equation in Eq.~(\ref{eqn:GK}) for passing particles can be written using an integrating factor \citep[Eqs.~(15)-(16)]{connor1980stability},
\begin{equation}
    g_p=-i\sigma(\omega-\tilde{\omega}_\star)f_0\int_{-\sigma\infty}^\ell \frac{J_0\phi}{|v_\parallel|}e^{i\sigma M(\ell',\ell)},
\end{equation}
and,
\begin{equation}
    M(a,b)=\int_a^b \frac{\omega-\tilde{\omega}_d}{|v_\parallel|}\mathrm{d}\ell,
\end{equation}
with $\sigma$ the sign of $v_\parallel$.
In writing the solution for $g_p$ we implemented the usual vanishing conditions for $\Im\{\omega\}>0$. We will be considering the contribution from passing ions, treating the electron response to be adiabatic. 
\par
As the function $M(\ell',\ell)$ in the exponent of the integrand scales like $\sim\omega/\omega_t$, the exponent is large. We may then approximate the whole integral integrating by parts \citep{bender2013advanced}, and keeping terms up to order $O(\epsilon)$. In doing so we assume that $\partial_\ell\phi\sim\phi/\epsilon$ and $\omega_d$ to be ordered like $\omega$ or smaller.
\par
Once expanded, we must then integrate $g_p$ over velocity space and apply the quasineutrality condition \cite[Eq.~(19a)]{connor1980stability}. Upon careful explicit evaluation of the integrals over $v_\parallel$ we obtain plasma dispersion functions (resonance coming purely from the velocity dependence of $\tilde{\omega}_d$), and we may write the resulting quasineutrality condition as \cite[Eqs.(35)-(37)]{connor1980stability} after some algebra. 
\par
To derive the much simpler looking eigenvalue equation in the fluid limit of papers such as \cite{plunk2014,zocco2016}, one must, in addition to assuming the smallness of the streaming contribution to the mode, also assume the smallness of the drift frequency $\omega_d/\omega\ll1$. This particular ordering allows us to ignore the drift frequency resonance that led to the appearence of plasma dispersion functions, as in the large argument limit, ignoring this resonance simply incurs in an exponentially small error. With this expansion, the integral over $x_\perp$ left to be done simply become standard, and we shall use the same notation as in the text, Eqs.~(\ref{eqn:weber_integrals}), for these integrals, $F_n(b)$. Keeping the leading order terms it can be shown then that one has,
\begin{subequations}
    \begin{equation}
        (1+\tau-\frac{\alpha}{T})\phi-B\partial_\ell\left[\frac{\beta}{BT}\partial_\ell\phi\right]=0,
    \end{equation}
    \begin{equation}
        \frac{\alpha}{T}=\Gamma_0\left[1-\frac{\omega_\star}{\omega}(1-\eta)-\frac{\omega_d\omega_\star}{2\omega^2}\right]- F_2(b)\left(\frac{\omega_\star^T}{\omega}+\frac{\omega_\star\omega_d}{2\omega^2}\right)-\frac{\omega_\star^T\omega_d}{2\omega^2}F_4(b),
    \end{equation}
    \begin{equation}
        \frac{\beta}{T}=\frac{v_{Ti}^2\omega_\star^T}{\omega^3}F_2(b).
    \end{equation}
\end{subequations}
As a note of caution, here we took $m=1$ for the mass. Putting these all together in a more succinct way, and taking for simplicity $\omega_\star=0$ but $\omega_\star^T\neq0$, for our problem in which $B$ is constant, and the only spatial dependence in the field is in the drift frequency, we may write
\begin{equation}
    \left[(1+\tau-\Gamma_0)-\frac{\omega_\star^T}{\omega}b(\Gamma_0-\Gamma_1)+\frac{\omega_\star^T\omega_d}{2\omega^2}F_4(b)\right]\phi-\frac{\omega_t^2\omega_\star^T}{2\omega^3}\partial_{\bar{\ell}}^2\phi=0,
\end{equation}
where we have defined the transit frequency $\omega_t$ using the thermal velocity of the ions (which this equation attempts to describe) and some reference parallel length scale. This full-FLR form of the fluid equation is consistent with the fluid asymptotic limit of the dispersion function, Eq.~(\ref{eqn:gen_b_fluid_eqn}), discussed in this paper. The small $b$ limit is precisely (upon relaxing $\omega_\star=0$) of the form employed in \cite{plunk2014, zocco2016}.

\section{Notation glossary}
In Table~\ref{tab:symbols_ref} we summarise the notation employed in the paper for reference. Many of the variables used are standard in gyrokinetics. 

\begin{table}
    \centering
    \begin{tabular}{cp{0.3\textwidth}c|cp{0.3\textwidth}c}
        $\ell$ & Length along the field line & &
        $\bar{\ell}$ & Normalised length along the field line & \\
        $\alpha$ & Straight field line field line label & &
        $\psi$ & $2\pi$ times the toroidal magnetic flux (surface label) &  \\
        $q_i$ & Ion charge &   &
        $T_i$ & Ion temperature  & \\
        $\bar{n}$ & Density & &
        $\rho_i$ & Ion Larmor radius & \\
        $v_{Ti}$ & Thermal speed of ions & &
        $\omega$ & Mode frequency & \\
        $\gamma$ & Mode growth rate & &
        $\bar{\omega}_d$ & Negative of the drift frequency at bottom of bad curvature well &       (\ref{eqn:def_wd_model}) \\
        $\omega_\star$ & Density gradient driven diamagnetic drift & (\ref{eqn:GK}) &
        $\omega_\star^T$ & Temperature gradient driven diamagnetic drift & \\
        $\eta$ & Ratio of temperature to density gradient & &
        $\tilde{\omega}_d$ & Drift frequency (with velocity dependence) & \\
        $\tilde{\omega}_\star$ & Diamagnetic frequency (with velocity dependence) & &
        $\omega_t$ & Transit frequency & (\ref{eqn:wt_def}) \\
        $\Lambda$ & Half width of bad curvature region & (\ref{eqn:def_wd_model}) &
        $v_\parallel$ & Parallel velocity & \\
        $x_\parallel$ & Normalised parallel velocity &  & $\sigma$ & Sign of $v_\parallel$ & \\
        $v_\perp$ & Perpendicular velocity & &
        $x_\perp$ & Normalised perpendicular velocity & \\
        $k_\alpha$ & Poloidal wavenumber & (\ref{eqn:GK}) &
        $k_\psi$ & Normal wavenumber & (\ref{eqn:GK}) \\
        $\mathbf{k}_\perp$ & Perpendicular wavevector &  &
        $b$ & Finite Larmor radius parameter & (\ref{eqn:GK}) \\
        $F_{0i}$ & Leading order ion Maxwellian distribution &  &
        $g$ & Non-adiabatic perturbed distribution function & \\
        $g_n$ & $n$-th Taylor-Gauss mode of $g$ & (\ref{eqn:g_exp_herm}) &
        $\phi$ & Electrostatic potential & \\
        $\phi_n$ & $n$-th Taylor-Gauss mode of $\phi$ & (\ref{eqn:g_exp_herm}) &
        $\tau$ & Ratio of ion to electron temperature & (\ref{eqn:QN_schrod}) \\
        $\lambda$ & Exponential envelope parameter & (\ref{eqn:g_exp_herm}) &
        $\zeta$ & Kinetic parameter & (\ref{eqn:zeta_def_or}) \\
        $\delta$ & Small drift ordering parameter & &
        $\epsilon$ & Mode localisation ordering parameter & (\ref{eqn:def_epsilon})
    \end{tabular}
    \caption{\textbf{Glossary of notation.} Table including the symbols employed throughout the paper and their informal meaning. }
    \label{tab:symbols_ref}
\end{table}

\bibliographystyle{jpp}

\bibliography{jpp-instructions}

\end{document}